\documentclass[10pt, 3p,fleqn]{elsarticle}

\makeatletter
\def\ps@pprintTitle{%
 \let\@oddhead\@empty
 \let\@evenhead\@empty
 \def\@oddfoot{\centerline{\thepage}}%
 \let\@evenfoot\@oddfoot}
\makeatother

\usepackage[utf8]{inputenc}
\usepackage{graphics}
\usepackage{graphicx}
\usepackage{amsmath,amssymb,esint}
\usepackage{lineno}
\usepackage{multirow}
\usepackage{subfig}
\usepackage{natbib}
\usepackage{eucal}
\usepackage{mathtools}
\usepackage{placeins}
\usepackage{enumerate}
\usepackage[usenames,dvipsnames]{xcolor}
\usepackage[colorlinks]{hyperref}
\usepackage{setspace}
\usepackage{makecell}
\usepackage{booktabs}
\usepackage{listings}
\biboptions{sort&compress}
\journal{Journal of Computational Physics}

\begin{document}

\begin{frontmatter}

\title{Reducing volume and shape errors in front tracking by divergence-preserving velocity interpolation and parabolic fit vertex positioning}

\author{Christian Gorges}
\author{Fabien Evrard}
\author{Berend van Wachem}
\author{Fabian Denner\corref{cor1}}

\address{Chair of Mechanical Process Engineering, Otto-von-Guericke-Universit\"{a}t Magdeburg,\\ Universit\"atsplatz 2, 39106 Magdeburg, Germany}

\cortext[cor1]{fabian.denner@ovgu.de}

\begin{abstract}
    Volume conservation and shape preservation are two well-known issues related to the advection and remeshing in front tracking. To address these issues, this paper proposes a divergence-preserving velocity interpolation method and a parabolic fit vertex positioning method for remeshing operations for three-dimensional front tracking. Errors in preserving the divergence of the velocity field when interpolating the velocity from the fluid mesh to the vertices of the triangles of the front are a primary reason for volume conservation errors when advecting the front. The proposed interpolation method preserves the discrete divergence of the fluid velocity by construction and is compared in this work with other known interpolation methods in divergence-free and non-divergence-free test cases, with respect to volume conservation and shape preservation of the front. The presented interpolation method conserves the volume and shape up to an order of magnitude better than the conventionally used interpolation methods and is within the range of higher order interpolation methods at lower computational cost. Additionally, the parabolic fit vertex positioning method for remeshing operations locally approximates the front with a smooth polynomial surface, improving volume conservation and shape preservation by an order of magnitude compared to conventional remeshing algorithms.
\end{abstract}

\begin{keyword}
    Front tracking \sep
    Velocity interpolation \sep 
    Parabolic fit \sep 
    Volume conservation \sep
    Remeshing \\~\\
    \textcopyright~2022. This manuscript version is made available under the CC-BY-NC-ND 4.0 license. \href{http://creativecommons.org/licenses/by-nc-nd/4.0/}{http://creativecommons.org/licenses/by-nc-nd/4.0/}
\end{keyword}

\end{frontmatter}

\section{Introduction}
\label{sec:Introduction}
The simulation of multiphase flows with deformable interfaces is a challenging task in the field of computational fluid dynamics. The numerical representation of the interface separating at least two immiscible fluids is associated with two particular difficulties: its accurate transport through space, as well as the appropriate modelling and computation of the force resulting from surface tension. These two fundamental aspects are the main reasons for the emergence of two categories of numerical methods representing such interfaces: interface-capturing (volume) \citep{Hirt1981, DeBar1974, Osher1988} and interface-tracking (surface) \citep{Tryggvason2001, Popinet1999, Unverdi1992, Glimm1998} methods. Both of these approaches are based on the one-fluid formulation, where the involved fluids are described by the same set of governing equations, only distinguished by their fluid properties (i.e. density and viscosity). Additionally, the surface tension force is included in the governing equations in the form of a volumetric momentum source term.

Interface-capturing methods rely on an implicit representation of the interface, based on an interface indicator function. This implicitness means that the explicit position and shape of the interface is not known and an accurate computation of the geometric properties (i.e. curvature and interface normal), which are required for the surface tension force, can be difficult. On the other hand, this implicitness makes topology changes (i.e. coalescence and break-up) relatively easy to handle, which is one of the major advantages of interface-capturing methods. The most common interface-capturing methods are the volume-of-fluid (VOF) method \citep{Hirt1981, DeBar1974, Noh1976} and the level-set method \citep{Osher1988, Osher2003, Sussman1994}.

In contrast to the first category of methods, interface-tracking methods represent the interface explicitly. One possible way to do so is to continuously adapt the fluid mesh as to match the interface position, which is called the moving-mesh method \citep{Quan2007, Tukovic2012}. The benefit of this method is an accurate representation of the jump condition of the fluid properties and the surface tension force at the interface. The continuous adaptation of the fluid mesh requires a complicated implementation and, especially for large interface movements, is computationally expensive and difficult to parallelise. An alternative interface-tracking method is the front tracking method (FT) \citep{Tryggvason2001, Popinet1999, Unverdi1992, Glimm1998}, which is used in the scope of this work.

In front tracking, the interface is represented by a discrete surface whose vertices are advected by the underlying flow velocity field. In two dimensions (2D), the surface becomes a line, of which the vertices are connected by linear curves or cubic splines \citep{Popinet1999}, whereas in three dimensions (3D) commonly a triangulated mesh is created \citep{Tryggvason2001, Tryggvason2011a}. The major advantage arising from the explicit representation of the interface is the straightforward computation of the normal vector at each point on the interface and, particularly, of the curvature of the interface for the calculation of the surface tension force. In the works of \citet{Tryggvason2011a} as well as \citet{Shin2005}, the surface tension force is evaluated as a triangle or vertex centered value by a Frenet-Element or Frenet-Vertex method \cite{Bi2021}. Further possible computational methods for the curvature are given, for instance, by \citet{Todd1986},  \citet{Meyer2003} as well as \citet{deSousa2004}.

Since only the discrete surface is advected in front tracking, the local fluid properties, such as density and viscosity, have to be re-evaluated at every time instance. This re-evaluation entails the computation of an indicator function which represents the volume fraction of the fluids in each fluid mesh cell. The construction of this indicator function can be accomplished by smoothing the gradient of the jump of the indicator function from the front mesh onto the fluid mesh and solving the resulting Poisson equation \cite{Tryggvason2011a, Tryggvason2001}. Another possibility is to geometrically compute the volume fraction within a fluid mesh cell using the exact position of the front \cite{Aulisa2003a, Popinet1999}.

Despite the advantage of an accurate and straightforward representation of the geometric properties and surface tension force, front tracking presents several limitations. Front tracking does not naturally handle topology changes of the interface and complex procedures are required to identify and enforce merging or break-up instances \cite{Shin2005, Tryggvason2011a}. 
Furthermore, remeshing algorithms must be employed to address the distortion and compaction of the front during its advection \cite{Tryggvason2011a, Tryggvason2001}. Otherwise, the resolution will deteriorate where the interface expands, or wrinkles may occur at points where the interface is compressed. Both problems lead to a poor representation of the interface, which, in turn, leads to errors in the conservation of volume, and shape, as well as in the calculation of the surface properties. Additionally, standard remeshing strategies themselves can cause errors in shape preservation and volume conservation. Coarsening algorithms, in particular, tend to alter volume conservation as well as shape preservation of the front. The Memoryless-Simplification algorithm of \citet{Lindstrom1998} aims at preserving the local volume during its edge collapsing process. \citet{deSousa2004} developed the volume conserving TSUR-3D algorithm for smoothing small undulations appearing where the front contracts. Nevertheless, there is still great potential in improving the remeshing algorithms to optimally represent the interface and to reduce volume conservation and shape preservation errors. For this reason, a novel vertex repositioning algorithm is presented in this paper, whereby the triangulated front, in the course of mesh refinement and coarsening operations, more accurately represents the properties of the interface.

The general exchange of information between the front mesh and the underlying fluid mesh and, in this context especially, the velocity interpolation is not trivial and also requires careful consideration. As a result of errors in the velocity interpolation and in the time integration, the advection itself is also not formally conservative \cite{Tryggvason2001}. Errors due to time integration can be reduced relatively easily by using higher order methods \cite{Tryggvason2011a} but nonetheless, in most front tracking publications, first- or second-order Euler methods are used \cite{Tryggvason2001, Unverdi1992, Popinet1999, deSousa2004, Hua2008, Muradoglu2008, Pivello2014, vanSintAnnaland2006, Tolle2020}. The works of \citet{Terashima2010} and  \citet{Dijkhuizen2010} are one of the few front tracking studies where it is explicitly mentioned that third- or fourth-order Runge-Kutta time integration schemes are used. Additionally, \citet{Dijkhuizen2010} showed brief comparative results between first-order Euler and fourth-order Runge-Kutta time integration. An alternative procedure to overcome volume conservation errors is to apply a correction step after the front advection, which includes a global volume conserving repositioning of the vertices of the triangles of the front in the normal direction every few time steps, as first introduced by \citet{Tryggvason2001}. This is nowadays a widely used and effective method, which however lacks a physical basis \cite{Tryggvason2001, Tryggvason2011a, Terashima2009, Takeuchi2020}.

The volume conservation and shape preservation errors induced by the velocity interpolation are partly due to the interpolated velocity at the vertices of the triangles of the front not naturally preserving the divergence of the fluid mesh velocity field. This means, for instance, that if the underlying velocity field is divergence free, the velocity interpolated to the front vertices is not divergence free \cite{Tryggvason2011a}. Although this problem is well known, there is no best practice for the general use of multivariate interpolation methods and the choice depends strongly on the properties of the input data and the desired properties of the interpolated result. In the scope of front tracking, the desired property of the interpolated velocity field is that, at best, it preserves the divergence of the fluid velocity field, to keep errors in advection and conservation small. Additionally, the interpolated velocities should remain bounded with respect to the fluid velocities. It is also undesirable for the interpolation method to be extremely computationally intensive, since advection in tandem with interpolation are the basic building blocks of front tracking. To achieve these desired conditions for velocity interpolation in front tracking as good as possible, we present a novel, divergence-preserving velocity interpolation in front tracking.

The simplest and fastest interpolation approach for velocity interpolation in front tracking is the nearest-neighbour method, which was used in the very first front tracking simulations \cite{Tryggvason2011a}. The velocity at a front vertex is simply adopted from the nearest fluid mesh point. This approach is generally too crude for an accurate advection and it is discontinuous, which is why it is no longer used and has been replaced by more accurate alternatives. Using the nearest 8 velocity points (in 3D) typically leads to the use of trilinear (volume-weighted) interpolation schemes, which are $C^0$ continuous and use first-order polynomials as interpolants. Because of its straightforward implementation and fast computation, this method is one of the most widely used interpolation methods in the field of front tracking \cite{Unverdi1992, Tryggvason2001, Popinet1999, vanSintAnnaland2006}. Another commonly used \cite[e.g.][]{deJesus2015,Hua2008,Muradoglu2008, Pivello2012, Pivello2014, Shin2018} interpolation method in this scope has been originally introduced by Peskin for the Immersed-Boundary method (IBM) \cite{Peskin1977, Peskin2003}. A Gaussian-like weighting kernel with a span of twice the fluid mesh spacing in every direction is superimposed on the target front vertex, resulting in a smooth interpolation of 64 mesh points (in 3D). Analogous to trilinear interpolation, triquadratic interpolation uses second-order polynomials and requires at least 27 sample points, whereas tricubic interpolation uses third-order polynomials as interpolants and requires a minimum of 64 sample points. Triquadratic and tricubic interpolation methods are not widely used for the velocity interpolation in front tracking, mostly owing to the (generally) second-order accuracy of the velocity field data. In contrast to trilinear, triquadratic and tricubic interpolation, the generally known spline interpolation methods interpolate piecewise between the specified sample points. Splines are usually twice continuously differentiable ($C^2$ continuous), which dampens oscillations as they can occur with standard polynomial interpolation due to unfavourable sample point positions. The most common spline is the cubic $C^2$ spline, but many other variants have been developed \cite{Engeln-Mullges2011}.

In addition to these well-known general-purpose schemes, interpolation methods have also been developed specifically to improve volume conservation and divergence preservation for interface tracking methods. Peskin's pioneering discoveries stand out in this field of research. In the work of \citet{Peskin1993}, a novel finite-difference divergence operator was developed, which is based on the arbitrarily chosen weighting kernel of the velocity interpolation method. This makes the discrete divergence at a mesh point the average of the continuous divergence within a box spanning around that mesh point. Because the underlying interpolation method serves as the basis for the new divergence operator, the volume conservation error for a divergence-free velocity field could be reduced by a factor of 67 \cite{Peskin1993}. Nevertheless, this method requires a far-reaching intervention in the discretisation of the underlying governing equations for the flow field, which proves tedious. \textcolor{black}{The parabolic edge reconstruction method (PERM) of McDermott and Pope \cite{McDermott2008}  uses a velocity reconstruction that is parabolic in the velocity-component direction and linear in the component-normal direction to preserve the discrete divergence of the velocity at the position of Lagrangian particles with second-order convergence. A major advantage of this method is that it is not only applicable to equidistant Cartesian meshes but also to non-uniform meshes. The divergence of the reconstructed field of PERM is piecewise trilinear in 3D, but may be discontinuous from cell to cell.}

In this study, we investigate the volume conservation and shape preservation errors caused by the velocity interpolation and remeshing in front tracking, and we propose a novel divergence-preserving velocity interpolation method and a vertex repositioning for remeshing algorithms based on a parabolic fit. Most of the front tracking methods described in the literature use first-order time integration schemes and linear or Peskin kernel interpolation, while only mentioning that higher-order time integration and other interpolation methods are easy to implement. This paper compares different interpolation methods with the proposed divergence-preserving interpolation method in terms of volume and shape errors. The focus is on local interpolations around the respective vertices of the triangles of the front and not on global interpolations of the entire velocity field in the vicinity of the front. In order to isolate the errors of interpolation and remeshing, artificial and analytically generated velocity fields are considered, in which the computation of surface tension force is not required. The new parabolic fit vertex repositioning for the common mesh coarsening and refining algorithms is presented and also analysed and compared in test cases for volume and shape errors. In a final step, the divergence-preserving interpolation method and the parabolic fit are applied to a rising bubble test case, as a reference for a real multiphase flow simulation, and the results are validated and compared to data of experimental and computational results reported in the literature.

This article is structured as follows: In Section \ref{section:VelocityInterpolation} we describe existing methods for the interpolation of the fluid velocity field on a fixed Cartesian fluid mesh to the vertices of the triangulated moving front mesh. We also propose the novel divergence-preserving velocity interpolation method to this means, and scrutinize this method by comparing its accuracy to existing methods for the two analytic test cases. In Section \ref{section:FrontRemeshing}, we propose the novel front vertex positioning method for remeshing operations based on a parabolic fit and compare its accuracy in terms of volume conservation and shape preservation to the classically used remeshing operations, where the vertex position is obtained without a parabolic fit. Finally, in Section \ref{section:RisingBubble}, we scrutinize the combination of the novel interpolation and remeshing methods using a realistic test case: rising bubbles at different conditions, and compare the accuracy with the current methods.

\section{Velocity interpolation}
\label{section:VelocityInterpolation}
In front tracking, the vertices of the discrete front are advected in a Lagrangian fashion by the equation
\begin{equation}
    \frac{\mathrm{d}\mathbf{x}_i (t)}{\mathrm{d}t} = \mathbf{u}(t, \mathbf{x}_i),
    \label{VertexAdvection1}
\end{equation}
where $\mathbf{x}_i$ is the position vector of the \textit{i}-th vertex of the front and $\mathbf{u}(\mathbf{x}_i)$ is the fluid velocity at $\mathbf{x}_i$. Because the fluid velocity is only known at discrete points on the fluid mesh, the velocity at the vertex position must be interpolated from the known mesh points. In practice, this leads to an advection equation at each point in time of the form
\begin{equation}
    \frac{\mathrm{d}\mathbf{x}_i}{\mathrm{d}t} = \mathbf{\bar{u}}(\mathbf{x}_i),
    \label{VertexAdvection2}
\end{equation}
where $\mathbf{\bar{u}}$ indicates an interpolated fluid velocity. The function under consideration or the discrete values are exactly reproduced by the interpolation function at the sample points, but only approximately in between them. This approximation leads to errors in the front advection and may also introduce divergence-related errors in the interpolated velocity field, even if the underlying flow field is divergence free.

In addition to these interpolation errors, the time discretisation of equation (\ref{VertexAdvection2}) may introduce additional errors in the advection of the front. The simplest approach is an explicit first-order Euler scheme, given as 
\begin{equation}
    \mathbf{x}_i\textcolor{black}{(t+\Delta t)} = \mathbf{x}_i\textcolor{black}{(t)} + \mathbf{\bar{u}}\textcolor{black}{(t,\mathbf{x}_i)} \Delta t + \mathcal O(\Delta t),
\end{equation}
which results in an error proportional to the time step $\Delta t$. First- and second-order Euler schemes are nowadays the most commonly used time discretisation schemes in front tracking \cite{Tryggvason2001, Unverdi1992, Popinet1999, deSousa2004, Hua2008, Muradoglu2008, Pivello2014, vanSintAnnaland2006, Tolle2020}. Using higher-order methods reduces the errors associated with the time discretisation in a straightforward way. For this reason, a classical fourth-order Runge-Kutta method is used in the further course of this study, resulting in
\begin{equation}
    \mathbf{x}_i\textcolor{black}{(t+\Delta t)} = \mathbf{x}_i\textcolor{black}{(t)} + \frac{\Delta t}{6} (k_1 + 2k_2 + 2k_3 + k_4) + \mathcal O(\Delta t^4),
\end{equation}
with
\begin{equation*}
    k_1 = \mathbf{\bar{u}}_i(t,\mathbf{x}_i),
\end{equation*}
\begin{equation*}
    k_2 = \mathbf{\bar{u}}_i(t + \frac{\Delta t}{2},\mathbf{x}_i + \frac{\Delta t}{2} k_1),
\end{equation*}
\begin{equation*}
    k_3 = \mathbf{\bar{u}}_i(t + \frac{\Delta t}{2},\mathbf{x}_i + \frac{\Delta t}{2} k_2),
\end{equation*}
\begin{equation*}
    k_4 = \mathbf{\bar{u}}_i(t + \Delta t,\mathbf{x}_i + \Delta t k_3).
\end{equation*}
\textcolor{black}{Since the fourth-order Runge-Kutta method requires four velocity evaluations per time step and the Euler method only one, the Runge-Kutta method is computationally more expensive, but for a given time step, the Runge-Kutta method yields a more accurate result compared to the Euler method. However, the cost of advecting the front is, overall, negligible in comparison with the cost of solving the equations governing the flow. Hence, the relatively low additional costs of the Runge-Kutta method for better accuracy of the front advection do not play a role for the overall computational effort.}

In the following, a second-order accurate divergence-preserving interpolation scheme based on the work of \citet{Toth2002} is proposed for the interpolation of the velocity from the fluid mesh to the vertices of the triangles of the front. The aim is to improve volume conservation and shape preservation in comparison to the velocity interpolation methods classically used in front tracking, with only a modest increase of computational costs. A brief introduction to the comparative interpolation methods is given after the new divergence-preserving method is explained in more detail. The fluid mesh is considered uniform and Cartesian throughout this study.

\subsection{Divergence-preserving interpolation}
\label{section:DivFreeIntp}
\begin{figure}[t]
\centering
\includegraphics[]{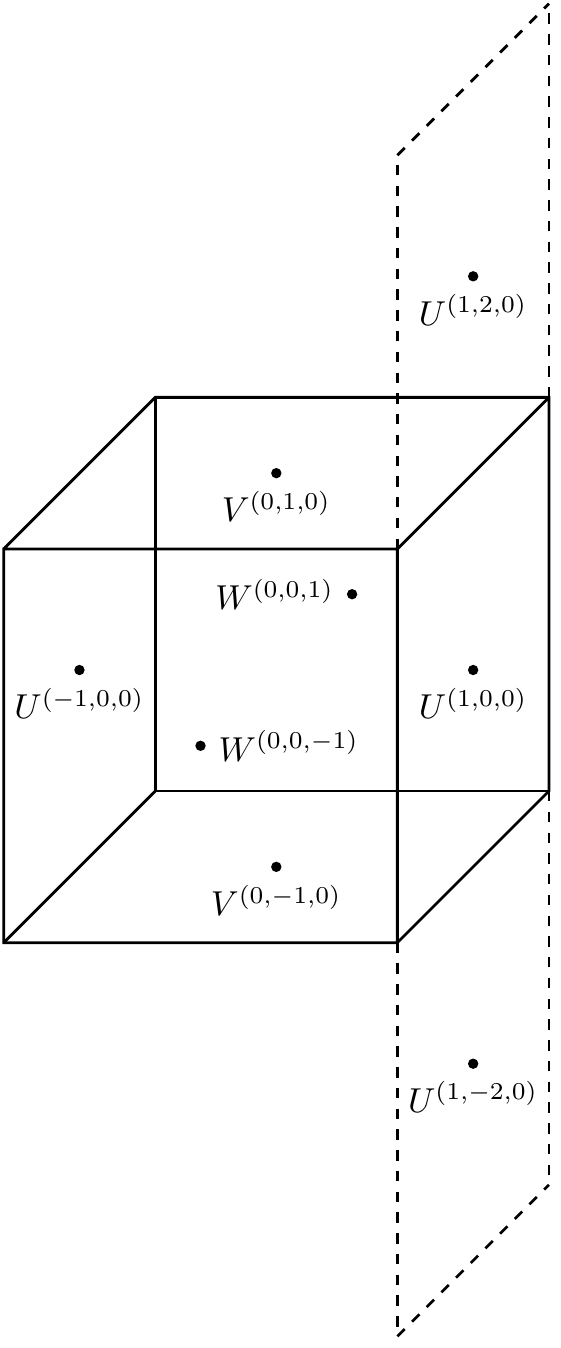}   
\caption{Positions of the face-centered values for the computation of the divergence-preserving velocity interpolation.}
\label{fig:DivfreeInt}
\end{figure}

\citet{Toth2002} proposed a second-order interpolation method for staggered 2D and 3D Cartesian meshes, which preserves the divergence of a discretized velocity field for the interpolated velocities. Originally, the method was developed for adaptive and hierarchical remeshing in the scope of magnetohydrodynamics, but can be used for any case that requires a divergence-preserving interpolation, such as for the velocity interpolation in front tracking. \textcolor{black}{In this work, the proposed divergence-preserving interpolation method is adopted without modifications from the work of \citet{Toth2002}.} For simplicity, a Cartesian 3D fluid mesh cell ranging from $\begin{bmatrix}-1 & -1 & -1 \end{bmatrix}$ to $\begin{bmatrix}1 & 1 & 1 \end{bmatrix}$ is considered, as illustrated in Figure \ref{fig:DivfreeInt}. The discrete normal face velocities at the cell faces are $U^{(\pm1,0,0)}$, $V^{(0,\pm1,0)}$ and $W^{(0,0,\pm1)}$, where the superscripts denote the coordinates of the face centers.
Given these discrete face-centered values the exact discrete divergence of the cell is given by:
\begin{equation}
\nabla \cdot {\mathbf{U}} = \frac{1}{2} \left( U^{(1,0,0)} - U^{(-1,0,0)} + V^{(0,1,0)} - V^{(0,-1,0)} + W^{(0,0,1)} - W^{(0,0,-1)}\right). 
\end{equation}
Furthermore, central-difference approximations of the transverse gradients $U^{(\pm1,0,0)}_y$, $U^{(\pm1,0,0)}_z$, $V^{(0,\pm1,0)}_x$, $V^{(0,\pm1,0)}_z$, $W^{(0,0,\pm1)}_x$ and $W^{(0,0,\pm1)}_y$ are defined by following the same reasoning, as exemplary shown for
\begin{equation}
    U^{(1,0,0)}_y = \frac{1}{4} \left(U^{(1,2,0)} - U^{(1,-2,0)} \right).
    \label{TransverseGradients}
\end{equation}
The linear interpolation of the face-centered velocities at any position $\begin{bmatrix}x & y & z \end{bmatrix}$ inside the cell yields
\begin{equation}
    \bar{\mathbf{u}}^{\prime}(x,y,z) = \begin{bmatrix} \bar{u}^{\prime}(x,y,z) \\ \bar{v}^{\prime}(x,y,z) \\ \bar{w}^{\prime}(x,y,z) \end{bmatrix},
\end{equation}
with
\begin{align}
    \bar{u}^{\prime}(x,y,z) & = 
    \frac{1+x}{2} \left(U^{(1,0,0)} + yU_y^{(1,0,0)} + zU_z^{(1,0,0)}\right) +
    \frac{1-x}{2} \left(U^{(-1,0,0)} + yU_y^{(-1,0,0)} + zU_z^{(-1,0,0)}\right) \label{eqLin1}\\
    \bar{v}^{\prime}(x,y,z) & = 
    \frac{1+y}{2} \left(V^{(0,1,0)} + xV_x^{(0,1,0)} + zV_z^{(0,1,0)}\right) +
    \frac{1-y}{2} \left(V^{(0,-1,0)} + xV_x^{(0,-1,0)} + zV_z^{(0,-1,0)}\right) \\
    \bar{w}^{\prime}(x,y,z) & = 
    \frac{1+z}{2} \left(W^{(0,0,1)} + xW_x^{(0,0,1)} + yW_y^{(0,0,1)}\right) +
    \frac{1-z}{2} \left(W^{(0,0,-1)} + xW_x^{(0,0,-1)} + yW_y^{(0,0,-1)}\right).
    \label{eqLin2}
\end{align}
The divergence of this linearly interpolated velocity inside the cell is
\begin{equation}
\begin{split}
    \nabla \cdot \bar{\mathbf{u}}^{\prime} 
    & = \frac{1}{2} \left[ \left(U^{(1,0,0)} + yU_y^{(1,0,0)} + zU_z^{(1,0,0)}\right) - \left(U^{(-1,0,0)} + yU_y^{(-1,0,0)} + zU_z^{(-1,0,0)}\right) \right.\\
    & \quad\quad\left. + \left(V^{(0,1,0)} + xV_x^{(0,1,0)} + zV_z^{(0,1,0)}\right) -  \left(V^{(0,-1,0)} + xV_x^{(0,-1,0)} + zV_z^{(0,-1,0)}\right) \right.\\
    & \quad\quad\left. + \left(W^{(0,0,1)} + xW_x^{(0,0,1)} + yW_y^{(0,0,1)}\right) -  \left(W^{(0,0,-1)} + xW_x^{(0,0,-1)} + yW_y^{(0,0,-1)}\right)\right],
\end{split}
\end{equation}
which differs from the discrete divergence $\nabla \cdot {\mathbf{U}}$ of the fluid mesh cell. To preserve the discrete divergence of the velocity field, corrections $f$, $g$, $h$ are introduced for each velocity component, such that inside the cell:
\begin{align}
    \bar{u}(x,y,z) & = \bar{u}^{\prime}(x,y,z) + f(x,y,z) \\
    \bar{v}(x,y,z) & = \bar{v}^{\prime}(x,y,z) + g(x,y,z) \\
    \bar{w}(x,y,z) & = \bar{w}^{\prime}(x,y,z) + h(x,y,z).
\end{align}
The derivation of the correction terms is given in \ref{Appendix A}. The correction terms to preserve the exact discrete divergence inside the cell are:
\begin{align}
    f(x,y,z) & = \frac{1-x^2}{4} \left(V_x^{(0,1,0)} - V_x^{(0,-1,0)} + W_x^{(0,0,1)} - W_x^{(0,0,-1)}\right) \label{eqcor1}\\
    g(x,y,z) & = \frac{1-y^2}{4} \left(U_y^{(1,0,0)} - U_y^{(-1,0,0)} + W_y^{(0,0,1)} - W_y^{(0,0,-1)}\right)\\
    h(x,y,z) & = \frac{1-z^2}{4} \left(U_z^{(1,0,0)} - U_z^{(-1,0,0)} + V_z^{(0,1,0)} - V_z^{(0,-1,0)}\right).
    \label{eqcor2}
\end{align}
The interpolated velocity $\mathbf{\bar{u}} = \begin{bmatrix}\bar{u} & \bar{v} & \bar{w} \end{bmatrix}$  at the position $\begin{bmatrix}x & y & z \end{bmatrix}$ inside an arbitrarily sized Cartesian cell ranging from $\begin{bmatrix}x_0 & y_0 & z_0 \end{bmatrix}$ to $\begin{bmatrix}x_1 & y_1 & z_1 \end{bmatrix}$ is given by the divergence-preserving interpolation polynomials as:
\begin{equation}
    \begin{split}
       \bar{u}(x,y,z) & = 
    \frac{x - x_0}{\Delta x} \left(U^{(+,0,0)} + \left( y - \frac{y_0 + y_1}{2} \right)U_y^{(+,0,0)} + \left( z - \frac{z_0 + z_1}{2} \right)U_z^{(+,0,0)}\right)\\ 
    &+ \frac{x_1 - x}{\Delta x} \left(U^{(-,0,0)} + \left( y - \frac{y_0 + y_1}{2} \right)U_y^{(-,0,0)} + \left( z - \frac{z_0 + z_1}{2} \right)U_z^{(-,0,0)}\right)\\
    &+ \frac{(x_1 - x)(x - x_0)}{2\Delta x} \left(V_x^{(0,+,0)} - V_x^{(0,-,0)} + W_x^{(0,0,+)} - W_x^{(0,0,-)}\right),
    \end{split}
\end{equation}
\begin{equation}
    \begin{split}
\bar{v}(x,y,z) &= \frac{y - y_0}{\Delta y} \left( V^{(0,+,0)} + \left( x - \frac{x_0 + x_1}{2} \right) V^{(0,+,0)}_x + \left( z - \frac{z_0 + z_1}{2} \right) V^{(0,+,0)}_z \right)\\\
    &+ \frac{y_1 - y}{\Delta y} \left( V^{(0,-,0)} + \left( x - \frac{x_0 + x_1}{2} \right) V^{(0,-,0)}_x + \left( z - \frac{z_0 + z_1}{2} \right) V^{(0,-,0)}_z \right)\\\
    &+ \frac{(y_1 - y)(y - y_0)}{2\Delta y} \left(U_y^{(+,0,0)} - U_y^{(-,0,0)} + W_y^{(0,0,+)} - W_y^{(0,0,-)}\right),
    \end{split}
\end{equation}
\begin{equation}
    \begin{split}
\bar{w}(x,y,z) &= \frac{z - z_0}{\Delta z} \left( W^{(0,0,+)} + \left( x - \frac{x_0 + x_1}{2} \right) W^{(0,0,+)}_x + \left( y - \frac{y_0 + y_1}{2} \right) W^{(0,0,+)}_y \right)\\\
    &+ \frac{z_1 - z}{\Delta z} \left( W^{(0,0,-)} + \left( x - \frac{x_0 + x_1}{2} \right) W^{(0,0,-)}_x + \left( y - \frac{y_0 + y_1}{2} \right) W^{(0,0,-)}_y \right)\\\
    &+ \frac{(z_1 - z)(z - z_0)}{2\Delta z} \left(U_z^{(+,0,0)} - U_z^{(-,0,0)} + V_z^{(0,+,0)} - V_z^{(0,-,0)}\right).
    \end{split}
\end{equation}
\textcolor{black}{The superscripts $(\pm,\pm,\pm)$ of the face centered velocities $U,V,W$ and their transverse gradients indicate the position of the corresponding faces with respect to the cell center.}
The first two terms on the right-hand side are the linear interpolation terms from equations (\ref{eqLin1})-(\ref{eqLin2}), whereas the last terms on each right-hand side are the correction functions of equations (\ref{eqcor1})-(\ref{eqcor2}). These interpolation functions ensure that the discrete divergence is preserved. The interpolation functions are continuous across cell faces in the normal direction, but they are discontinuous in the transverse direction, because the transverse gradients do not exactly match on both sides of a face connecting two cells \cite{Toth2002}. The main issue with an interpolated velocity field being discontinuous is that, as a front vertex crosses the face between two cells, its trajectory may be subject to a discontinuity as the velocity changes abruptly. This typically leads to oscillations. Nonetheless, the discontinuity vanishes with increasing fluid mesh resolution with second order \cite{Toth2002}.

\subsection{Polynomial interpolation}
Polynomial interpolation is the search for a polynomial that runs exactly through a given set of points. For $n+1$ given pairs of values $(x_i,y_i)$ with different interpolation points $x_i$, a polynomial $P$ of the $n$-th degree is sought that satisfies all the equations
\begin{equation}
    P(x_i) = y_i, \text{ with }i=\{0,\ldots,n\}.
\label{Polynominterpolation}
\end{equation}
Such a polynomial always exists and is uniquely determined. Equation (\ref{Polynominterpolation}) leads to a linear system of equations. The \textit{trilinear interpolation} is the easiest and one of the most common multivariate polynomial interpolation methods for velocity interpolation in front tracking on 3D fluid (Cartesian) meshes \cite{Unverdi1992, Tryggvason2001, Popinet1999, vanSintAnnaland2006}. It uses first-order polynomials to converge with second order accuracy for decreasing sample point distances. The main reasons for its popularity are a straightforward implementation and fast computation, because only 8 sample velocities are required for the interpolation. A simplified and fast approach, which yields the same results as trilinear interpolation, is the use of weighting kernels consisting of a product of one dimensional functions for the 8 nodal values of a 3D fluid mesh cell:
\begin{equation}
    \bar{\mathbf{u}}_i(x_i,y_i,z_i) = \sum_{l=1}^{8} \mathbf{u}(\mathbf{x}_l) w(\mathbf{x}_l).
    \end{equation}
The subscript $l$ indicates the considered sample points, which are in this case the 8 nodal velocities of the mesh cell. The weight of a sample point is given as
\begin{equation}
    w(x_l,y_l,z_l) = d(r_x)d(r_y)d(r_z),  
\end{equation}
with
\begin{equation}
    r_x = \frac{(x_i - x_l)}{\Delta x},
\end{equation}
where $\Delta x$ is the mesh cell spacing. The values for $r_y$ and $r_z$ are computed in an analogous fashion. For completion, $d(r)$ follows as
\begin{equation}
    d(r) = \begin{cases}
    1 - r, & 0 \leq r < 1 \\
    1 + r, & -1 < r \leq 0 \\
    0, & |r| \geq 1. \\
    \end{cases}
\end{equation}
This approach is referred to as \textit{volume-weighted interpolation}, because the weights can be interpreted as volume fractions of the cell \cite{Tryggvason2011a}.

The second interpolation method considered in this study is \textit{triquadratic interpolation}, which is also a classical polynomial interpolation, but of second-order polynomials. The linear equation system resulting from the triquadratic polynomial,
\begin{equation}
   P(\mathbf{x}) = \sum_{i=0}^2 \sum_{j=0}^2 \sum_{k=0}^2 a_{i,j,k} \mathbf{x}_{i,j,k},
\end{equation}
can be solved efficiently using at least 27 sample points and converges with third order \cite{Engeln-Mullges2011}. The interpolation quality does not usually improve when more sample points are used. Especially with equidistant points and with increasing degree of the polynomial, strong oscillations may occur and convergence is not necessarily given for polynomial interpolation (Runge's phenomenon) \cite{Engeln-Mullges2011}.

\subsection{Peskin interpolation}
The standard interpolation schemes Peskin introduced for the immersed boundary method \cite{Peskin2003,Peskin1972} are still one of the mostly used velocity interpolation methods in the field of front tracking \cite{deJesus2015,Hua2008, Pivello2012, Pivello2014}. For the sake of simplicity, these approaches are henceforth called \textit{Peskin interpolation}. In this work, the Peskin weighting kernel 
\begin{equation}
    d(r) = \begin{cases}
    \frac{1}{4} \left( 1 + \cos \left( \frac{1}{2} \pi r \right) \right), & |r| < 2 \\
    0, & |r| \geq 2 \\
    \end{cases}
    \label{PeskinEquation}
\end{equation}
with a span of twice the fluid mesh spacing in every direction is used. It results in a smoothed interpolation of 64 mesh velocity points around the target point:
\begin{equation}
    \bar{\mathbf{u}}_i(x_i,y_i,z_i) = \sum_{l=1}^{64} \mathbf{u}(\mathbf{x}_l) w(\mathbf{x}_l),
\end{equation}
with $w(\mathbf{x}_l) = d(r_x) d(r_y) d(r_z)$, where $r_x = (x_i - x_l)/\Delta x$.
\subsection{Spline interpolation}
While polynomial interpolation can oscillate up to the point of unusable interpolation results due to improperly positioned sample points, spline interpolation always delivers usable interpolation results. Spline interpolation attempts to interpolate given sample points, also called knots or interpolation points, with the help of piecewise polynomials in between these knots. Since spline interpolation always converges, it usually provides a lower order of convergence than polynomial interpolation. Within the field of spline interpolation there are many variants. One of these is the so-called \textit{thin-plate-spline interpolation} which is the generalisation of one-dimensional natural cubic splines \cite{Engeln-Mullges2011} to higher dimensions. The spline surface represents a thin-plate sheet that is constrained not to move at the sample points. The construction is based on radial basis functions that exactly interpolate the sample points, minimising an integral that represents the bending energy of the surface. The origin and a detailed derivation of thin-plate-splines is given in \cite{DuchonJean1976,DuchonJean1977} and the extension to higher dimensions is explained in \cite{MeinguetJean1979}.

While the classical natural cubic spline is twice continuously differentiable ($C^2$ continuous), the \textit{Akima (cubic) spline} is only once continuously differentiable ($C^1$ continuous) \cite{Akima1970, Engeln-Mullges2011}. An important property of the Akima spline is its locality in the construction of the coefficients of the interpolation polynomial between any two knots. This means that there is no large system of equations to solve. Compared to natural cubic splines, the Akima spline produces fewer undulations and is better suited to deal with quick changes between flat regions and converges with second order. In this study, the implementations of the thin-plate-spline and Akima spline are provided by the Geometric Tools Engine (GTE) mathematics library \cite{Eberly2021} and both are applied to a $3 \times 3 \times 3$ stencil of sample velocity points.

\subsection{Numerical tests}
\label{TestsInterpolation}
The proposed divergence-preserving interpolation method is tested for two test cases, with a divergence-free as well as a non-divergence-free velocity field. \textcolor{black}{The other interpolation methods discussed above serve as comparative benchmarks in terms of the shape, vertex position and volume errors of the front.}

\textcolor{black}{The volume of a body enclosed by the triangulated mesh representing the discrete front can be computed as the sum of the volumes between each mesh triangle and its projection onto the $x=0$ plane. The volume between the $i$-th triangle and its projection is given as
\begin{equation}
    V_i = \frac{1}{6} N_x (v_{1,x} + v_{2,x} + v_{3,x}),
    \label{Volumenberechnung}
\end{equation}
where $N_x$ is the $x$-component of the cross-product of vectors defined by $\mathbf{v}_1 - \mathbf{v}_0$ and $\mathbf{v}_2 - \mathbf{v}_0$ and $v_{i,x}$, with $i=1,2,3$, are the $x$ components of the triangle vertices $\mathbf{v}_i$. Subsequently, the relative volume error follows as
\begin{equation}
    \epsilon_V = \frac{V_{\mathrm{int}} - V_{\mathrm{ref}}}{V_{\mathrm{ref}}},
    \label{volumenfehlerformel}
\end{equation}
where $V_{\mathrm{int}}$ refers to the volume enclosed by the front, when advected with interpolated velocities, and $V_{\mathrm{ref}}$ is the reference volume enclosed by the front, when advected with the exact, analytical velocity (i.e. without discrete interpolation).}
 
As a proxy for the shape error, the relative error of the front vertex norm or the relative radius error, both averaged over all vertices of the triangles of the front, are calculated as
\begin{equation}
    \epsilon_P = \frac{\| \mathbf{x}_{\mathrm{int}} \| - \| \mathbf{x}_{\mathrm{ref}} \| }{ \| \mathbf{x}_{\mathrm{ref}} \|},
    \label{PositionError}
\end{equation}
\begin{equation}
    \epsilon_R = \frac{R_{\mathrm{int}} - R_{\mathrm{ref}}}{R_{\mathrm{ref}}},
    \label{RadiusError}
\end{equation}
where the subscript $\mathrm{int}$ refers to the value of the advected front with velocity interpolation. \textcolor{black}{The reference values $\mathbf{x}_\text{ref}$ are the positions of the front vertices that are advected with the exact, analytical velocity.} The radius is $R = \| \mathbf{x} - \mathbf{x}_c \|$, where the subscript $c$ indicates the position of the center of the volume bounded by the interface.

In the following, the fourth-order Runge-Kutta time integration method is used for all simulations.

\subsubsection{Expanding sphere}
\begin{figure}[ht]
    \centering
    \includegraphics{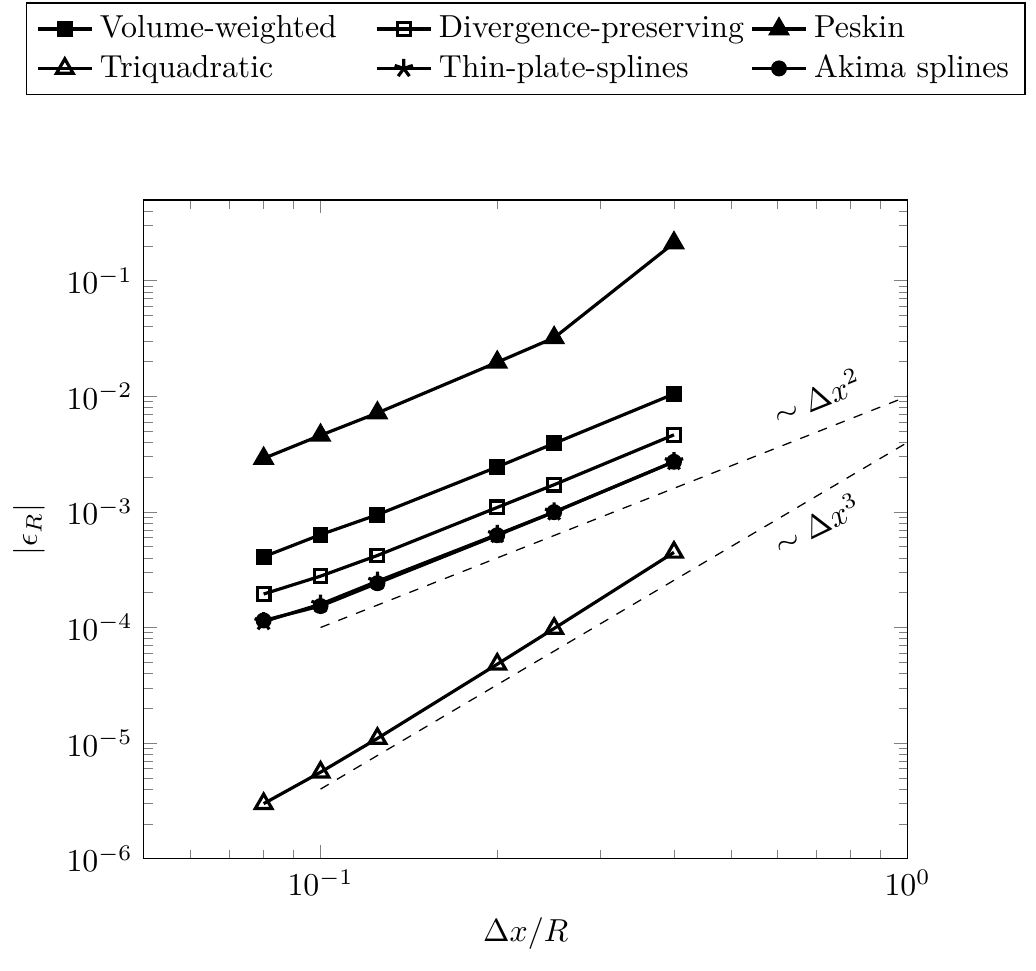}
    \caption{Relative radius errors of the considered interpolation methods for the expanding sphere test case. The sample velocities for interpolation are the exact velocities at the face centers of the fluid mesh.}
    \label{fig:Velocity_Int_ExpSphere_ExactNodalVel}
\end{figure}

The first test case is a sphere that expands in the radial direction with the velocity field prescribed as
\begin{equation}
\mathbf{u}(\mathbf{x}) = U \frac{\mathbf{x}}{\| \mathbf{x} \|} ,
\label{expandingspherevelocityfield}
\end{equation}
where $U = \| \mathbf{u} \|$ is the reference velocity. The velocity field, defined on the Cartesian fluid mesh with spacing $\Delta x$, is pointing outwards from the center and is not divergence free. The sphere with an initial radius $R=0.1$ is represented by a front mesh with 1280 triangles (the initial triangle edge length to radius ratio $l/R$ is $\approx 0.151$). The CFL-number, defined as $\mathrm{CFL} = U \Delta t / \Delta x$, is kept constant at approximately $0.12$. After the sphere has expanded for $t=3 R/U$, the average relative radius error is computed as $\Delta x/R$ varies. Figure \ref{fig:Velocity_Int_ExpSphere_ExactNodalVel} shows the results of the relative radius errors obtained by interpolating the exact velocity from the face centers of the fluid mesh to the vertices of the triangles of the front using different interpolation methods. Each interpolation method achieves the expected order of convergence. The triquadratic method exhibits third-order convergence, whereas the proposed divergence-preserving and the remaining methods converge with second order. Both spline methods and the triquadratic interpolation have a smaller absolute error compared to the other methods. The largest stencil of 64 sample velocity points for the Peskin interpolation seems to smooth the velocity field too much, producing the largest absolute error throughout. The proposed divergence-preserving interpolation method produces errors on the same order of magnitude as the spline methods and is more accurate than the volume-weighted (trilinear) and Peskin interpolation methods.

\begin{figure}[ht]
    \centering
    \includegraphics{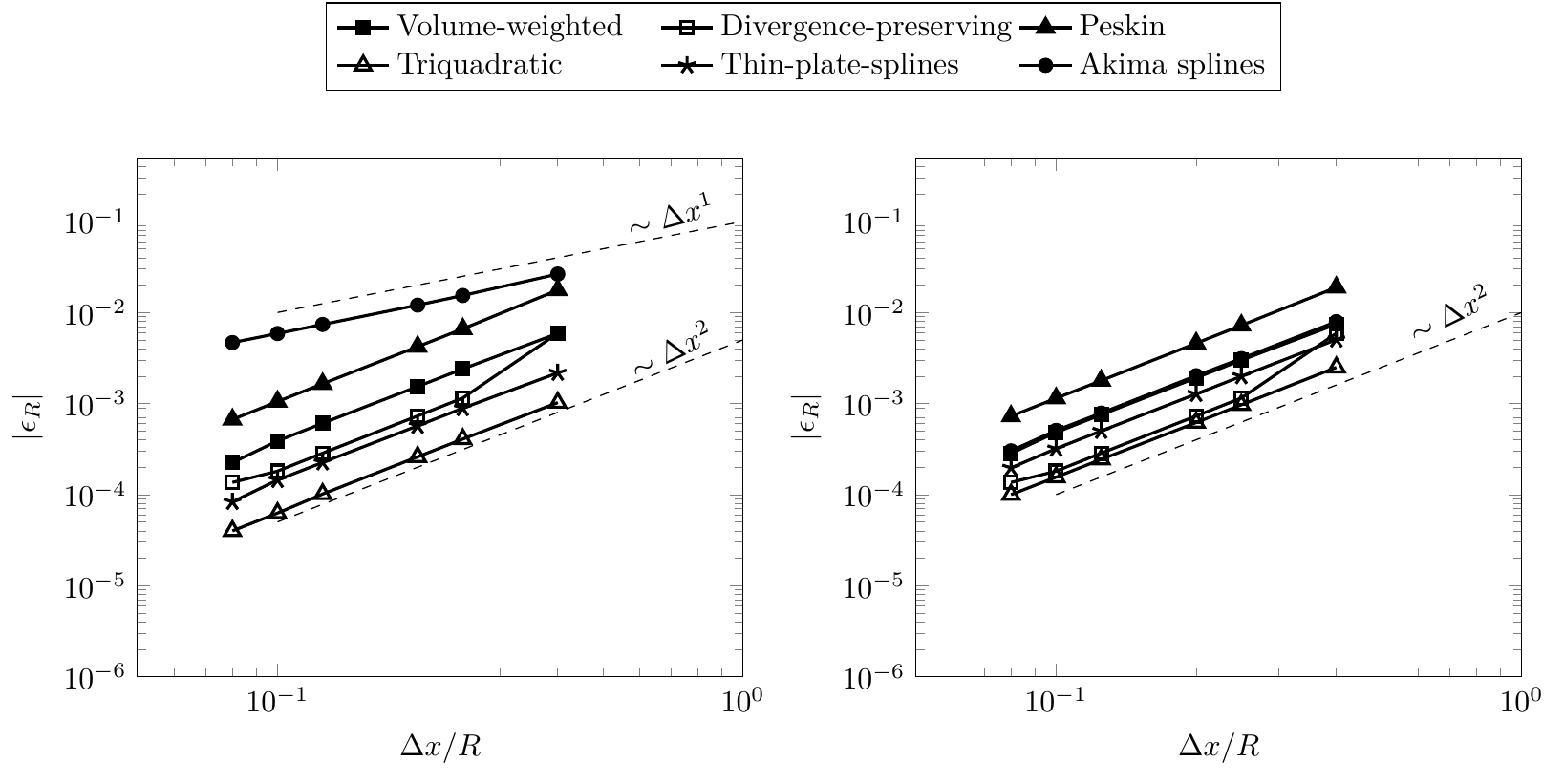}
    \caption{Radius error of the considered interpolation methods for the expanding sphere test case. The left diagram represents a staggered finite-volume framework, where the sample velocities for interpolation are second-order approximations of the exact velocity at the face centers of the fluid mesh. The right diagram represents a collocated finite-volume framework, where the sample velocities are second-order approximations of the exact velocity at the cell centers of the fluid mesh.}
    \label{fig:Velocity_Int_ExpSphere_FVFramework_Both}
\end{figure}

However, in a finite-volume framework, the velocities at the face or cell centers are usually second-order approximations. 
Thus, the velocity field within a standard finite-volume framework is, by construction, only second-order accurate and interpolating such a velocity field with higher-order interpolation methods will not improve the accuracy. For this reason, the interpolation methods are evaluated again with face-centered as well as cell-centered velocities approximated with second-order accuracy. The results are shown in Figure \ref{fig:Velocity_Int_ExpSphere_FVFramework_Both}. The left diagram shows the results for the face-centered velocities and the left diagram for the cell-centered velocities. It should be noted that the divergence-preserving interpolation method is only defined on the basis of the face-centred velocities and the results of this method in the right diagram have only been copied from the left diagram for better comparison. All interpolation methods are now converging with second order for both types of finite-volume frameworks, except the Akima spline interpolation which deteriorates to first order for the radius error in the face-centered velocity interpolation case, but is also second-order accurate for the cell-centered velocity interpolation case. In general, for both second-order approximation approaches, the absolute errors of all methods are closer to each other compared to the interpolation of the exact velocities, with the proposed divergence-preserving method still being more accurate than the volume-weighted (trilinear) and the Peskin interpolation, and close to the triquadratic interpolation method.

\subsubsection{Interface deformation}
\label{Analytic deformation velocity field}
\begin{figure}[ht]
    \centering
    \includegraphics[width=1.0\textwidth]{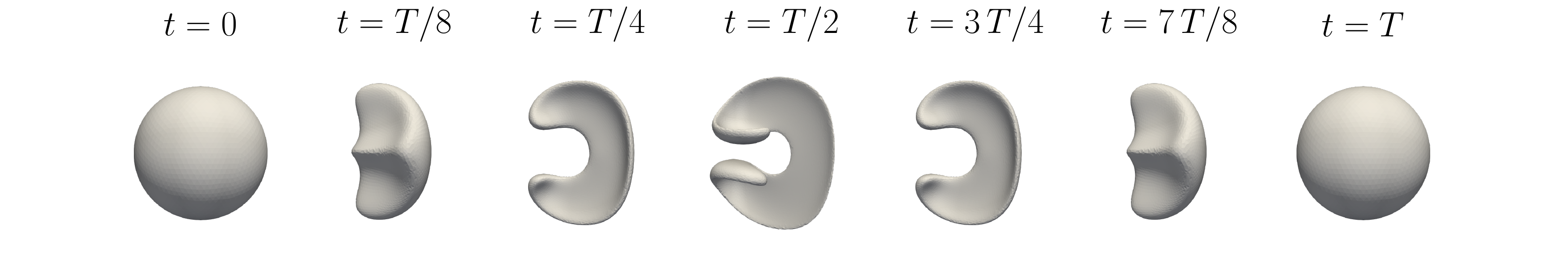}
    \caption{Evolution (from left to right) of the sphere in the deforming velocity field of \citet{LeVeque1996}. Up to $t=T/2$ the initially spherical front gets stretched and afterwards (under ideal conditions) returns to initial position and shape.}
    \label{fig:LeVequeFig}
\end{figure}
The second test case for the comparison of the different interpolation methods and the validation of the proposed divergence-preserving interpolation method is the three-dimensional interface deformation in a divergence-free velocity field of \citet{LeVeque1996}. The initially spherical interface is deformed by the analytically defined velocity field, given as
\begin{equation}
\begin{split}
    &u(\mathbf{x}, t) = 2 \sin^2 \left(\pi x \right) \sin \left( 2\pi y \right) \sin \left( 2\pi z \right) \cos \left( \pi t/T \right)\\\
    &v(\mathbf{x}, t) = -\sin\left( 2\pi x \right) \sin^2 \left( \pi y \right) \sin \left( 2\pi z \right) \cos \left( \pi t/T \right)\\\
    &w(\mathbf{x}, t) = -\sin \left( 2\pi x \right) \sin \left( 2\pi y \right) \sin^2 \left( \pi z \right) \cos \left( \pi t/T \right).\\
\end{split}
\end{equation}
The velocity field is time dependent and applied on the time interval $0\leq t \leq T$, where $T$ is set {\it a priori}. Until $t=T/2$ the front gets quite strongly deformed and from $t=T/2$ the velocity field is reversed, such that, under ideal conditions, the initial position of the sphere is recovered at $t=T$. The errors introduced by the numerical methods are the only mechanisms that can cause differences between the start and end positions of the front. \textcolor{black}{In order to consider the shape errors caused by the interpolation method, these are evaluated at $t=T/2$ rather than at $t=T$. The main reason for this is that the velocity field in this test case is symmetric, in time, with respect to $T/2$. This means that some of the advection errors introduced between the start and $T/2$ will be cancelled during the return phase from $T/2$ to $T$. To remove this bias in the interpretation of the accuracy of the interpolation methods, we choose to evaluate the errors at $T/2$.} The way in which the interface is deformed depends largely on its initial position. Figure \ref{fig:LeVequeFig} shows the evolution of the deforming front initially centered at $\begin{bmatrix}0.35 & 0.35 & 0.35 \end{bmatrix}$ with radius $R=0.15$ inside a unit cube for $T = 3.0$.
The reference velocity of this velocity field is $U=2$ and the CFL-number is kept constant at approximately $0.24$ for various fluid mesh spacings.

\begin{figure}[ht]
    \centering
    \includegraphics{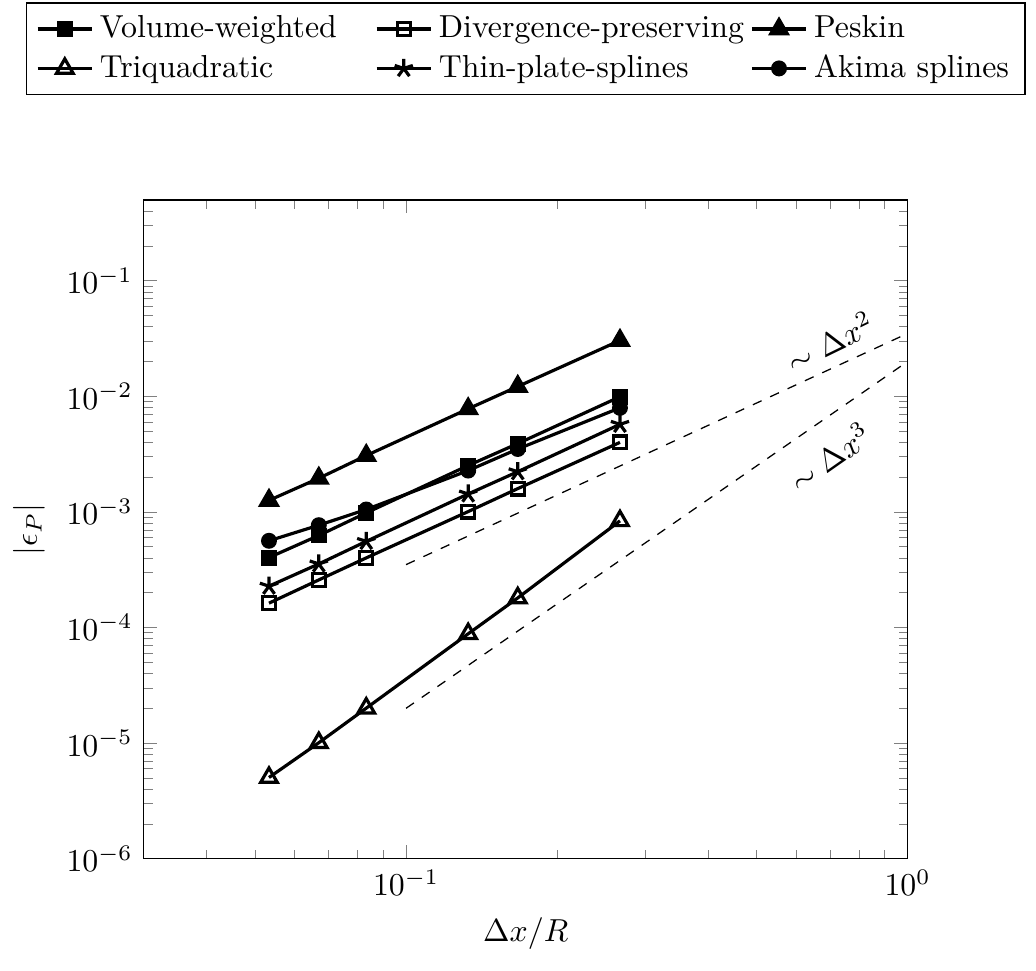}
    \caption{The error norm of the front vertex position of the considered interpolation methods for the interface deformation test case. The sample velocities for interpolation are the exact velocities at the face centers of the fluid mesh.}
    \label{fig:Velocity_Int_LeVeque_ExactNodalVel}
\end{figure}

In Figure \ref{fig:Velocity_Int_LeVeque_ExactNodalVel} the results for the interpolation of the exact velocities at the face centers are shown. Compared to the previously shown expanding sphere, only the absolute errors of the individual methods are different, the order of convergence remains the same for all methods. The proposed divergence-preserving interpolation method is slightly more accurate than the thin-plate-splines in this case and only the triquadratic interpolation achieves even more accurate results. The Peskin and Akima interpolation methods show the least accurate results, followed by volume-weighted (trilinear) interpolation.
Considering the results for the second-order approximated face-centered velocities in the left diagram of Figure \ref{fig:Velocity_Int_LeVeque_FVFramework_Both} and the cell-centered velocities in the right diagram of Figure \ref{fig:Velocity_Int_LeVeque_FVFramework_Both} (the results for the divergence-preserving interpolation method are still based on face-centered velocities because of its construction), the proposed divergence-preserving interpolation method exhibits clearly the smallest errors. The error norm of the front vertex position is approximately one order of magnitude smaller with the proposed divergence-preserving interpolation method compared to the second best performing method, the triquadratic interpolation.

\textcolor{black}{Furthermore, the volume error is calculated at time $T/2$, which is shown for the different interpolation methods using the face-centered velocities in Figure \ref{fig:Velocity_Int_LeVeque_Verror}. The ranking of the volume errors for the different interpolation methods follows the same ordering as for the vertex norm error in Figure \ref{fig:Velocity_Int_LeVeque_FVFramework_Both}. The divergence-preserving velocity interpolation method yields the smallest volume error among the considered interpolation methods.}

\textcolor{black}{
\begin{figure}[ht]
    \centering
    \includegraphics{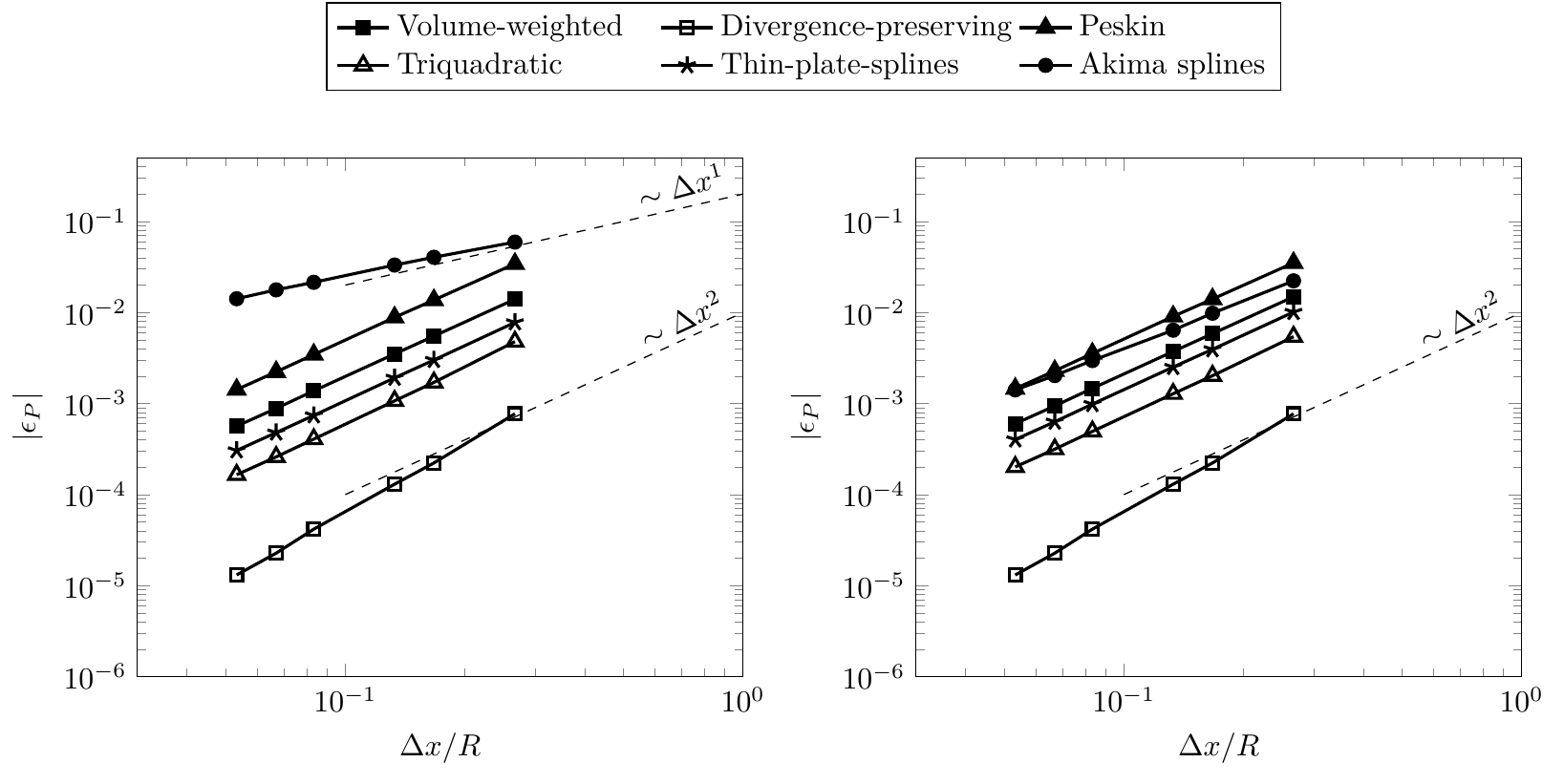}
    \caption{\textcolor{black}{The error norm of the front vertex position of the considered interpolation methods for the interface deformation test case. The sample velocities for each interpolation method in the left diagram are face-centered velocities, as they are readily available in a staggered finite-volume framework. The sample velocities for each interpolation method (except the divergence-preserving interpolation method, which is only defined for face-centered sample points) in the right diagram are cell-centered velocities, as they are readily available in a collocated finite-volume framework. Both, the face-centered and the cell-centered velocities are second-order approximations of the exact velocities at the respective face- or cell-centered sample point. To achieve this, the exact velocities of the analytic flow field are integrated over the area of the respective face or the volume of the respective cell.}}
    \label{fig:Velocity_Int_LeVeque_FVFramework_Both}
\end{figure}
\begin{figure}[ht]
    \centering
    \includegraphics{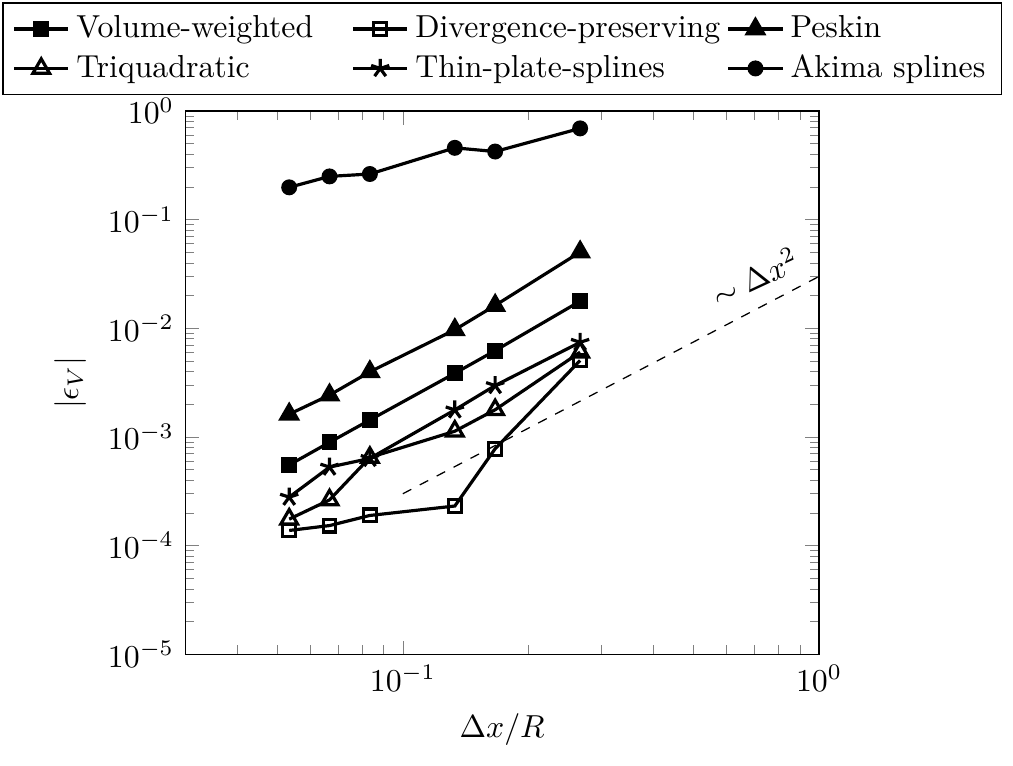}
    \caption{\textcolor{black}{Volume error of the considered interpolation methods for the interface deformation test case. The sample velocities for interpolation are second-order approximations of the exact velocity at the face centers of the fluid mesh.}}
    \label{fig:Velocity_Int_LeVeque_Verror}
\end{figure}
}

\section{Front remeshing}
\label{section:FrontRemeshing}
As a result of the fluid flow, the fluid interface is moved, compressed and stretched. Initially, evenly distributed vertices of the triangles of the front are, thus, distributed unevenly across the front while the interface evolves. For instance, compression of the front crowds the front vertices, whereas stretching the front pulls the vertices apart. At first glance, increased front vertex or triangle density may not appear to be a problem, other than an unnecessary increase in computational overhead. However, small non-physical undulations may occur at these locations, due to variations in the velocity field and an inaccurate vertex advection \cite{deSousa2004, Tryggvason2001}. A reduction in front vertex density due to stretching of the front results in an under-resolved front mesh. An under-resolved front mesh leads to a poor representation of the interface, with a loss of information that cannot be recovered later on, and in an inaccurate determination of the geometric properties, such as the interface normal vector and curvature. Consequently, this leads to an inaccurate calculation of the surface tension force, which may cause artificial pressure perturbations and spurious currents \cite{Tryggvason2011a}. For these reasons, it is essential to continuously adapt the front mesh during its evolution.

\begin{figure}[ht]
    \centering
    \includegraphics{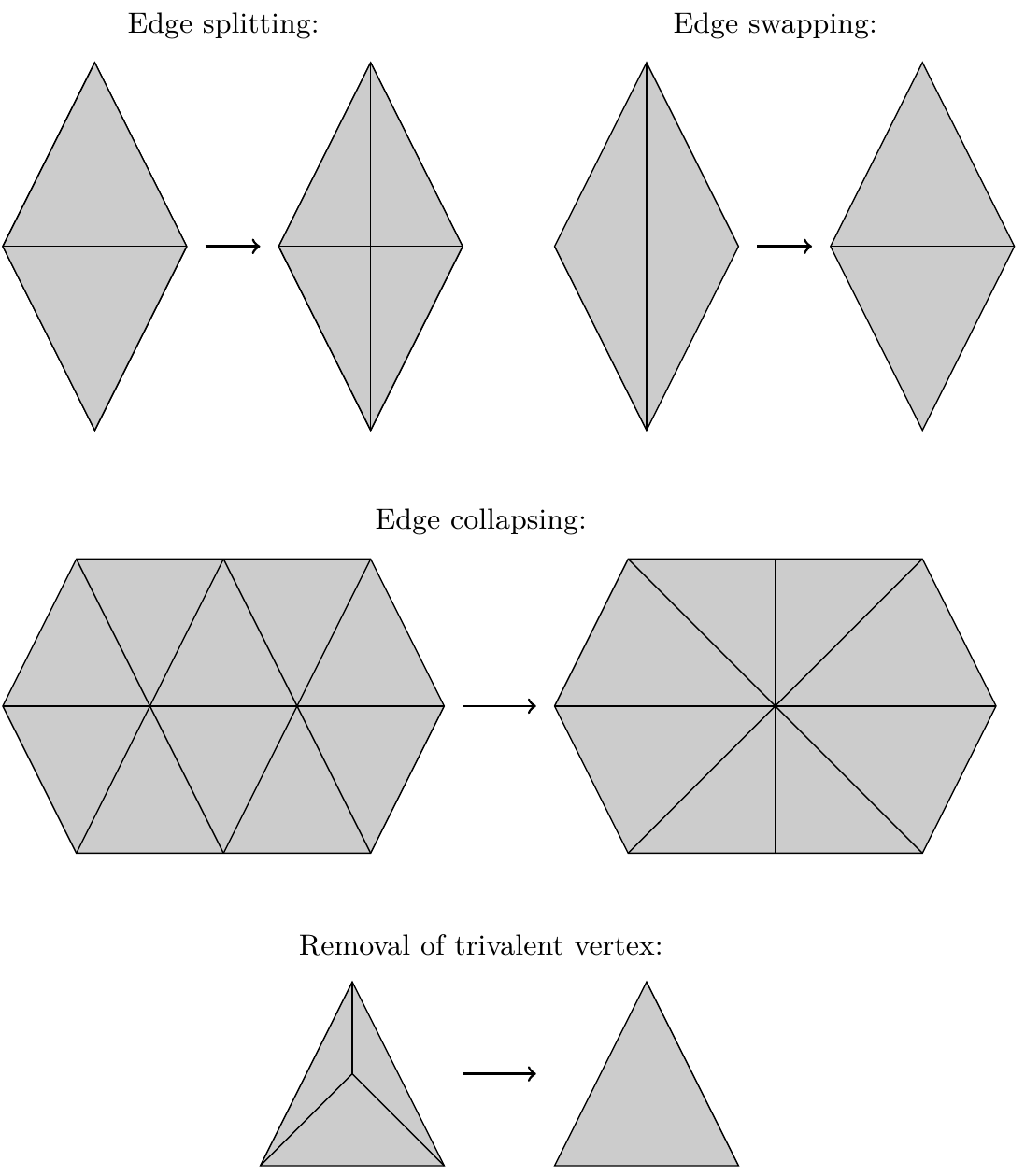}
    \caption{Remeshing operations for a 3D triangulated front mesh.}
    \label{fig:RemeshingOperations3DFronttracking}
\end{figure}

For front meshes in 3D front tracking simulations, remeshing operations are most commonly based on edge manipulation \cite{Tryggvason2001, Tryggvason2011a, Pivello2014, Pivello2012, Hua2008}. The remeshing operations are illustrated schematically in Figure \ref{fig:RemeshingOperations3DFronttracking}. 
The mesh refinement is achieved by increasing the vertex or triangle density. For this purpose, selected edges are split by adding a new vertex. Additionally, the splitting of an edge results in two new triangle elements. The opposite, mesh coarsening, causes an edge to collapse, removing one vertex and two triangles. In addition to these two main operations, there are other methods that are not mandatory for refining or coarsening the mesh, but may improve the quality of the mesh in some circumstances. Examples for operations applied to improve the mesh quality include edge swapping and removing trivalent vertices, also illustrated in Figure \ref{fig:RemeshingOperations3DFronttracking}. 

\subsection{Remeshing operations}
When the remeshing operations are based on edge manipulation, the main criterion is usually the edge length. If the edge length is above a predefined threshold value the edge will be split, if the edge is shorter than a predefined threshold it will be collapsed. Further remeshing criteria are, for instance, the triangle element size or its aspect ratio \cite{Tryggvason2011a}. For an accurate computation and communication of variables, for instance volume fraction and surface tension force, from the front to the fluid mesh it is preferable to ensure that there is at least one front vertex or triangle element center (depending on the computational method for the surface tension force) inside a fluid mesh cell that contains a part of the interface. Therefore, the edge length thresholds for splitting and collapsing should be somehow based on the fluid mesh spacing. \citet{Tryggvason2011a} proposed an edge length of $\Delta x/3 \leq l_e \leq \Delta x $, where $\Delta x$ is the fluid mesh spacing and $l_e$ is the edge length of the front mesh. The criteria for edge splitting and collapsing in this work are based on the criteria from \citet{Jiao2010} and are listed in Table \ref{tab:EdgeSplittingCollapsingCriteria}. The splitting criteria refine triangles that are too large or poorly shaped, with large angles and long opposite edges. The first two collapsing criteria in Table \ref{tab:EdgeSplittingCollapsingCriteria} address poorly shaped triangles with a very small angle or a short edge, whereas the third and fourth condition remove triangles that are well shaped but too small \cite{Jiao2010}. The interaction of all these criteria is complex and the parameters must be chosen so that splitting and collapsing are consistent with each other. In addition, the edge length of the front triangle must still match the mesh spacing of the fluid mesh. The general relations of the threshold parameters in Table \ref{tab:EdgeSplittingCollapsingCriteria} are $s<S<l<L$, $\theta_s \ll \theta_l$ and $r<R$, where $s, S, l, L$ are the length thresholds, $\theta_s, \theta_l$ are angle thresholds and $r, R$ are ratios for calculating the length thresholds based on the desired edge length $l$. Furthermore, \citet{Jiao2010} proposed that $s \approx 0.27l$, $S=R l$, $L=1.5l$, $\theta_s \approx 15^{\circ}$ and $\theta_l \approx 145^{\circ}$ with $r=0.25$ and $R=0.45$. To be in some way consistent with Tryggvason's \cite{Tryggvason2011a} suggestion for front tracking simulations, the desired edge length $l$ should be approximately $0.5 \Delta x \leq l \leq \Delta x$. These threshold values are used in the scope of this work, but, nevertheless, these values are only suggestions and optimal setting parameters may vary from case to case.

\begin{table}[ht]
    \centering
     \caption{Edge splitting and collapsing criteria based on \citet{Jiao2010}. }
    \label{tab:EdgeSplittingCollapsingCriteria}
    \begin{tabular}{ll}
    \toprule
     & \makecell[c]{Edge splitting}\\
    \midrule
    \makecell[l]{Absolute long edge} & \makecell[l]{The edge is longest among those of its incident triangles\\ and longer than threshold $L$.}\\
    \midrule
    \makecell[l]{Relative long edge} & \makecell[l]{The edge is longer than the desired length $l$ ($l<L$),\\ one of its opposite angles is greater than the threshold $\theta_l$\\ and shortest edge of the incident triangles is not shorter than $s$ ($s<l$).}\\
    \bottomrule\\[-0.5em]
    % \end{tabular}
    
    % \begin{tabular}{cc}
    \toprule
     & \makecell[c]{Edge collapsing}\\
    \midrule
    \makecell[l]{Absolute small angle} & \makecell[l]{The opposite angle in an incident triangle is smaller than the threshold $\theta_s$\\ and longest edge of the triangle is shorter than the desired length $l$.}  \\
    \midrule
    \makecell[l]{Relative short edge} & \makecell[l]{The edge is shorter than the fraction $r$ of the longest edge in its incident triangles.}  \\
    \midrule
    \makecell[l]{Absolute small triangle} & \makecell[l]{The longest edge in the incident triangles is shorter than the threshold $S$\\ and the considered edge is the shortest.} \\
    \midrule
    \makecell[l]{Relative small triangle} & \makecell[l]{The longest edge in the incident triangles is shorter than $l$\\ and the considered edge is shorter than the fraction $R$ of the longest edge.}  \\
    \bottomrule
    \end{tabular}
\end{table}

In addition to these individual operation criteria, it is important to ensure that all remeshing operations do not negatively influence the topology of the front mesh. The standard edge splitting operation, whereby the new vertex is positioned at the center of the considered edge, does not affect local volume conservation, but may misrepresent the local interface curvature. On the other hand, edge collapsing replaces an edge by a vertex, and therefore has the potential to change the local volume, the shape of the front and the mesh quality. If a collapsed edge is just simply substituted by its midpoint, the front will locally contract in convex regions and inflate in concave regions. For this reason, edges in strongly curved areas are not collapsed. Emblematic for the local curvature, the angle between the normal vectors of the adjacent triangles of the considered edge is computed and the edge is only collapsed if the angle is smaller than a threshold ($\approx 30^{\circ}$ throughout this work). The same is true for edge flipping.

The positioning of the new vertex in edge splitting and collapsing is still an issue for the conservation of volume and the preservation of the shape of the front. \textcolor{black}{In \cite{Tryggvason2011a}, Tryggvason et al. introduced a quadratic interpolation function in the barycentric coordinates for any type of vertex positioning, which requires six neighbouring vertices and is computationally cheap. The barycentric coordinate system is constructed locally by laying down grid lines that go through the vertices of the triangle under consideration and accounts for the local curvature.} \citet{Lindstrom1998} proposed an edge collapsing algorithm with volume-conserving vertex positioning, which has been used a few times in front tracking remeshing algorithms \cite{Pivello2012,Pivello2014}. Vertex positioning is treated as an optimization problem and the triangle quality, the local volume and the local area are constraints limiting the space for calculating the new coordinates. The volume conservation is based on calculating the volumes of tetrahedons under the affected triangles. This method is not applicable to edge splitting, since the volume is not changed by this operation if only the center of the edge is considered as new vertex position. Additionally, this method may alter the local curvature and shape through the collapsing operation. For these reasons, in the following section, we present a vertex repositioning method that is applicable to both edge splitting and collapsing.

\subsection{Parabolic fit vertex positioning}
To improve the conservation of volume and the preservation of shape through edge splitting and edge collapsing remeshing operations, the proposed method locally approximates the front mesh around the edge as a paraboloid, which is in the further course referred to as parabolic fit. Fitting a paraboloid to a sample of $n$ points $\mathbf{x}_i$ is, for example, also used in the context of curvature evaluation and surface reconstruction \cite{Ohtake2004a, Ohtake2004, Cazals2003, Evrard2017, Evrard2019}.
\begin{figure}[ht]
    \includegraphics[]{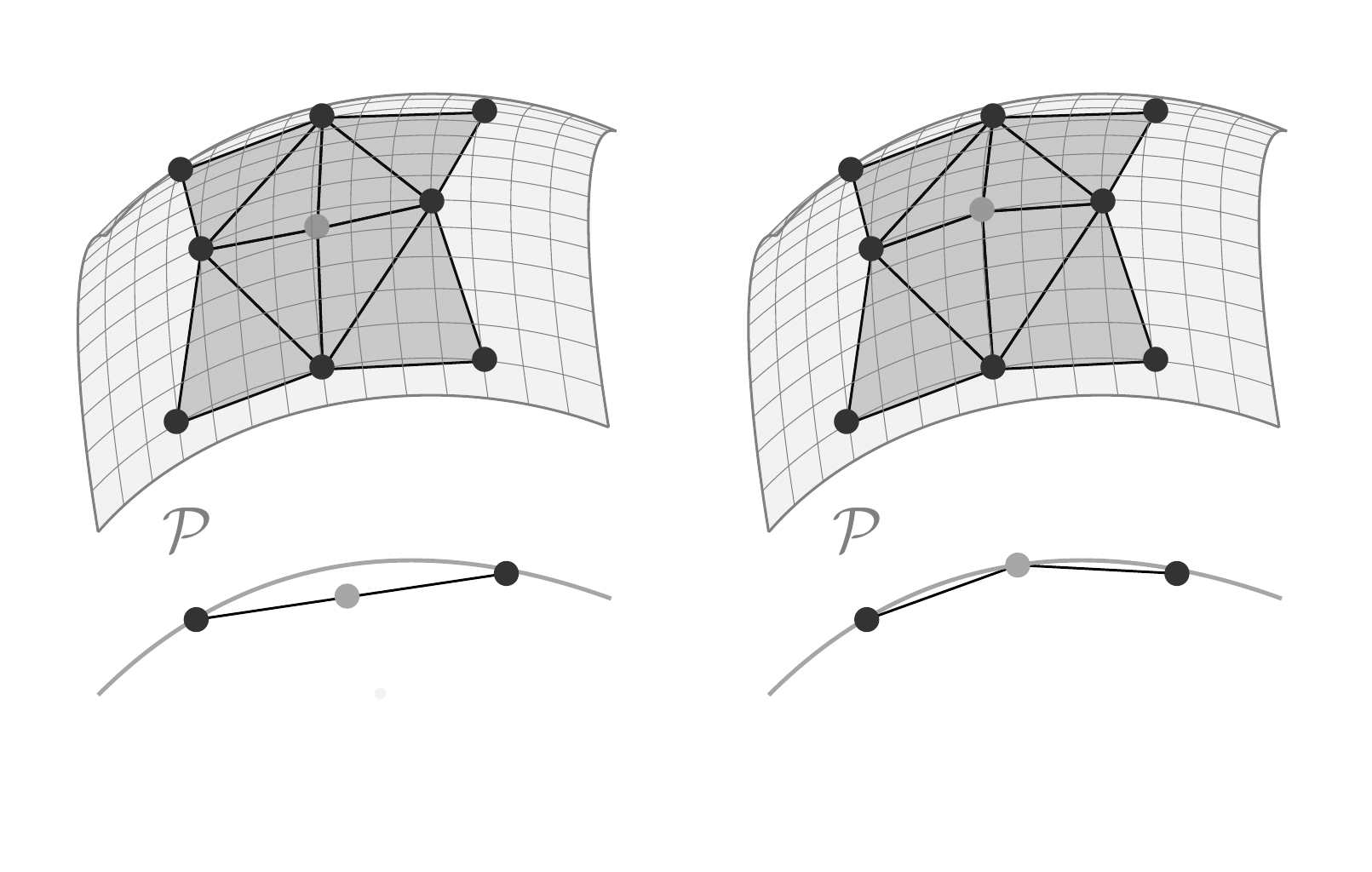}
    \caption{Edge split without (left) and with (right) parabolic fit. The implicit parabolic surface $\mathcal{P}$ is fitted to the direct neighbourhood vertices (the black vertices) and the mid-vertex (the grey vertex) is fitted to the parabolic surface. The adaptation to the mid-vertex of an edge to be collapsed is straightforward.}
    \label{RefiningWithFitInfluenceExample}
\end{figure}
First, a new local orthonormal basis $\begin{bmatrix}\mathbf{t}_1 & \mathbf{t}_2 & \mathbf{n} \end{bmatrix}$ is introduced whose origin $\mathbf{x}_c$ is at the midpoint of the edge under consideration (the grey point in Figure \ref{RefiningWithFitInfluenceExample}). The vector $\mathbf{n}$ is chosen as the average of the sample vertex normals and $\mathbf{t}_1 , \mathbf{t}_2$ are two tangential vectors arbitrarily chosen in the orthogonal plane. A comparison of various vertex normal evaluation methods is, for instance, given in \cite{Jin2005}. Next, the position vectors of the sample points (vertices of the triangles of the front) of the mesh (the black points in Figure \ref{RefiningWithFitInfluenceExample}) are transformed into the new local coordinate system: 
\begin{equation}
    \mathbf{x}_i^* = \mathbf{\mathcal{T}}^{-1} \left( \mathbf{x}_i - \mathbf{x}_c\right),
\end{equation}
where $\mathbf{\mathcal{T}} = \begin{bmatrix}\mathbf{t}_1 & \mathbf{t}_2 & \mathbf{n} \end{bmatrix}$ is the transformation matrix from the local to the global coordinate system.
In the local coordinate system, the implicit paraboloid surface can be defined as
\begin{equation}
    z = \mathcal{P}(x,y) = a_1 x^2 + a_2 x y + a_3 y^2 + a_4 x + a_5 y + a_6,
    \label{ParaboloidEq}
\end{equation}
with $\mathbf{x}^*=\begin{bmatrix}x & y & z \end{bmatrix}$.
The procedure of finding the six coefficients $\mathbf{a}=\begin{bmatrix}a_1 & a_2 & a_3 & a_4 & a_5 & a_6\end{bmatrix}$ can be described as a least squares problem finding the minimum algebraic distance between the implicit surface and the $n$ sample points $\mathbf{x}^*_i$  from the immediate vicinity of the edge to be split or collapsed:
\begin{equation}
    \text{min} \sum_{i=1}^n (z_i - \mathcal{P}(x_i , y_i))^2 .
    \label{LSAlgebraicfit}
\end{equation}
In matrix notation, the paraboloid $\mathcal{P}$ can be described as
\begin{equation}
    \mathcal{P}(x,y) = \mathbf{q}(x,y) \mathbf{a},
\end{equation}
where $\mathbf{q}$ is the polynomial basis function vector and $\mathbf{a}$ is the parameter vector. Finally, the solution to this least squares problem in equation (\ref{LSAlgebraicfit}) is given by solving
\begin{equation}
   \textcolor{black}{ \mathbf{a} = \left( \mathbf{Q}^{\text{T}} \mathbf{Q} \right)^{-1} \mathbf{Q}^{\text{T}} \mathbf{z},} 
\end{equation}
with $\mathbf{Q}$ the $n \times 6$ matrix defined by $Q_{ij} = q_j(x_i,y_i)$ and $\mathbf{z}$ the vector of the $z$ values. The new $z$-position of the mid-vertex of the considered edge in the local referential, which is shifted in the interface normal direction, is found by inserting the six coefficients into equation (\ref{ParaboloidEq}):
\begin{equation}
    z_m = a_1 x_m^2 + a_2 x_m y_m + a_3 y_m^2 + a_4 x_m + a_5 y_m + a_6,
\end{equation}
with $\mathbf{x}_m^* = \begin{bmatrix}x_m & y_m & z_m\end{bmatrix}$ the coordinates of the mid-vertex of the considered edge in the local referential. The new position in the original, global system is given as:
\begin{equation}
    \mathbf{x}_m = \mathbf{\mathcal{T}} \mathbf{x}_m^* + \mathbf{x}_c. 
\end{equation}
In order to be well posed, the fitting problem in equation (\ref{LSAlgebraicfit}) requires at least six independent sample vertices from the front mesh directly around the considered edge. Using more sample points than required does not usually improve the results, since the paraboloid should fit as closely as possible to the immediate neighbourhood and not be smoothed over a larger span. According to preliminary tests, using six to ten uniformly distributed sample vertices usually works best.

\textcolor{black}{It is possible to pose the paraboloid fitting problem as a constrained optimization problem for discretely conserving the volume encompassed by the front, at least for edge collapses. However, a vertex positioning strictly based on the conservation of volume would result in a poor positioning with regard to the preservation of the shape and curvature of the interface. Additionally, a strict local conservation of volume for vertex positioning in the edge-splitting algorithm would simply result in the use of the midpoint of the split edge. In summary, the user must decide whether volume conservation or shape conservation is the preferred property of the vertex positioning method. In the present study, we opted for a more accurate shape preservation in order to be able to calculate the geometric properties more precisely.}

\textcolor{black}{The least-squares problem in equation (\ref{LSAlgebraicfit}) can be solved using a commonly available least-squares solver (e.g. the function \textit{dgelsy} from the LAPACK-library).}

\subsection{Numerical tests}
In the following, the new parabolic fit repositioning method for edge splitting and collapsing is tested for volume and shape errors, using the test cases already introduced in section \ref{TestsInterpolation}. The quantities used for the error calculation are the volume enclosed by the front, the radius or position norm averaged over all vertices of the triangles of the front, and the curvature averaged over all triangle elements and its standard deviation.

The computation of the curvature at a triangle center is based on the parabolic fit procedure. For curvature computation, the center $\mathbf{x}_c$ of the local coordinate system is positioned at the center of the considered triangle. The curvature of the implicit parabolic height function at the triangle center is given by \cite{Evrard2020,Goldman2005}
\begin{equation}
    \kappa = \frac{\mathcal{P}_{xx} \left( 1+\mathcal{P}_{y}^2 \right) + \mathcal{P}_{yy} \left( 1+\mathcal{P}_{x}^2 \right) - 2 \mathcal{P}_{x}\mathcal{P}_{y}\mathcal{P}_{xy}}{\left( \mathcal{P}_{x}^2 + \mathcal{P}_{y}^2 +1 \right)^{\frac{3}{2}}},
\end{equation}
where the subscripts of $\mathcal{P}$ denote its partial derivatives.
The relative error for the vertex-averaged radius and the triangle-averaged curvature follow the same reasoning as the volume error calculation (equation (\ref{volumenfehlerformel})).

\subsubsection{Expanding and shrinking sphere}
The negative influence of front mesh coarsening operations on volume conservation and shape preservation is well known \cite{Pivello2014, Lindstrom1998}. However, the influence of edge splitting on volume conservation and shape preservation has so far remained largely unnoticed. For this reason, the following first test case explicitly considers the influence of edge splitting with or without parabolic fit of the mid-vertex of the split edge. \textcolor{black}{Additionally, the interpolation in the barycentric coordinates \cite{Tryggvason2011a} serves as a comparison.} A sphere with an initial radius of $0.1$ is expanded in the radial direction by the velocity field of equation (\ref{expandingspherevelocityfield}) until it reaches a radius of $R = 0.3$ at $t=4 R/U$. Then the velocity field is reversed and the sphere shrinks to its initial radius. The vertices of the triangles of the front are advected with the fourth-order Runge-Kutta method and the time step is $\Delta t = 0.01 R/U$. To single out the influence of the remeshing, the exact velocities at the vertices of the triangles of the front are applied. During the expansion of the sphere, the front mesh is refined exclusively by edge splitting. During the contraction phase, the front mesh remains at the refinement level reached at the end of the expansion phase, so no mesh coarsening is performed to explicitly highlight the influence of mid-vertex repositioning with or without parabolic fit at the end position. The desired front mesh edge length as basic remeshing threshold is $l = 0.0151$, which is the average edge length at the start of the simulation ($l/R = 0.151$). Every fifth time step, the remeshing criteria and thresholds are checked and, if necessary, the remeshing operations (in this case only edge splitting) are executed. \textcolor{black}{The frequency with which the remeshing criteria and thresholds should be checked is case dependent, and preliminary tests indicated that conducting these checks every fifth time step is sufficiently frequent for this particular case.} The geometric properties of the sphere at the start time serve as reference values for the error calculation of volume, radius and curvature at the end of the simulation, since the sphere would reach exactly the same position if numerical errors were absent.

\begin{figure}[ht]
\hspace{-2.0cm}
  \includegraphics[scale=0.9]{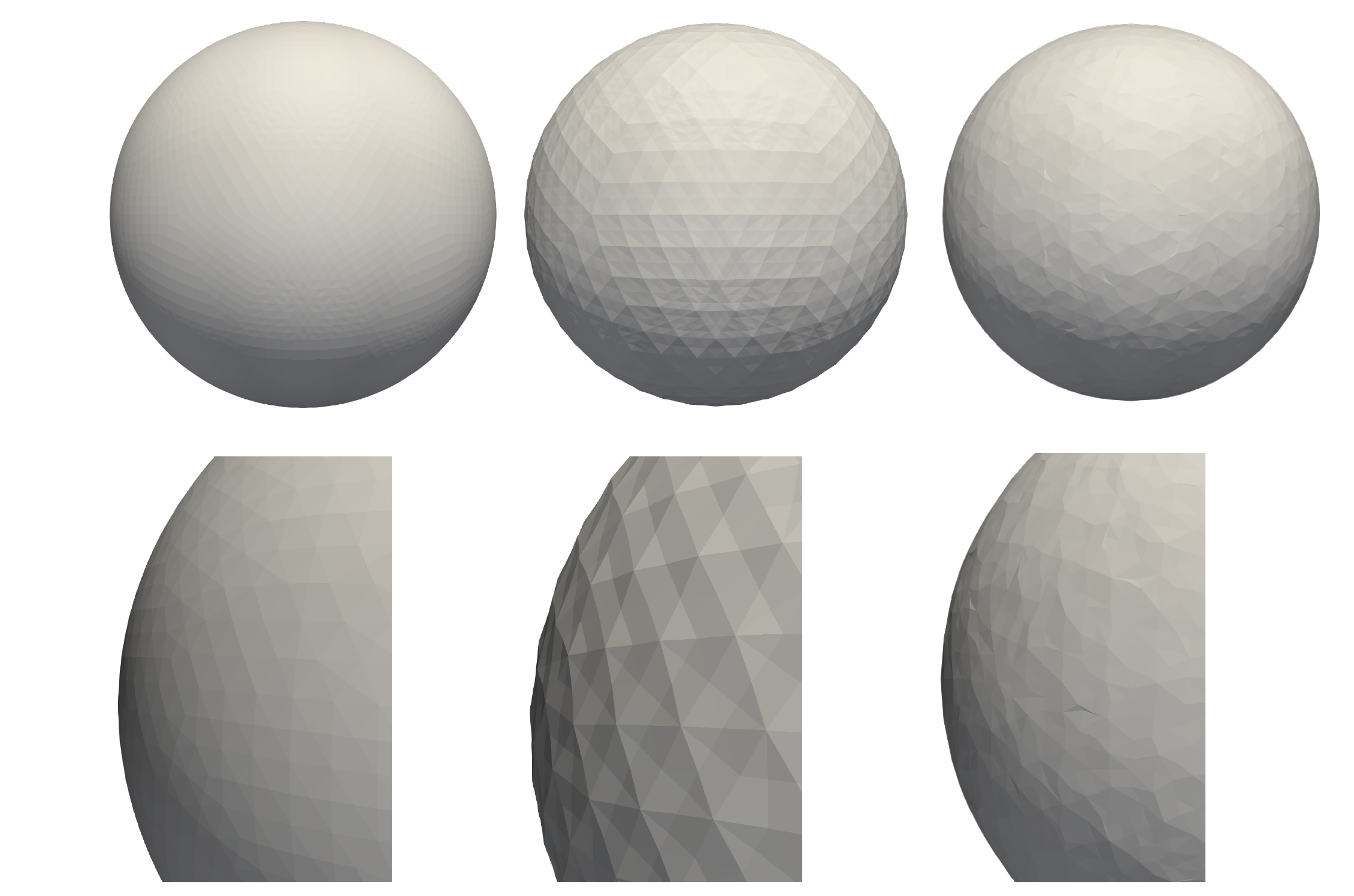}
    \caption{Surface mesh of a contracted front which has been refined using the parabolic fit during edge splitting (left images), without using the parabolic fit \textcolor{black}{(center images) and using the interpolation in the barycentric coordinates (right images).}}
    \label{SplittingWithAndWithoutFit}
\end{figure}

The results for edge splitting with or without parabolic fit, \textcolor{black}{and the interpolation in the barycentric coordinates,} for this test case are listed in Table \ref{SplittingOnlyResults}. The volume error for the simulation with parabolic fit shows an increase in volume at the end of the simulation compared to the start volume. This is to be expected, since the parabolic fit generates volume in convex regions to fit the front closer to the exact volume and shape of the interface. \textcolor{black}{The same is true for the interpolation in the barycentric coordinates, which has a slightly smaller volume error compared to the parabolic fit.} On the other hand, the sphere obtained without parabolic fit loses volume during the simulation. During edge splitting when the sphere expands, the volume remains constant because edge splitting without mid-vertex repositioning is volume conserving. Nevertheless, these newly generated vertices have, for a closed front, a smaller radius or distance to the center of the front. As the front contracts, this smaller distance becomes increasingly apparent in terms of a loss of volume and, in particular, as small undulations on the front. The phenomenon of the undulations is visible in Figure \ref{SplittingWithAndWithoutFit}, where the front of a sphere where the edges are split with the parabolic fit, which is smooth, is shown in the left images, and the front that occurs when a refined mesh without a parabolic fit or \textcolor{black}{barycentric interpolation for vertex positioning is contracted, which exhibits distinct undulations, are shown in the center images and the right images, respectively.} However, the absolute value of the volume error is slightly lower for the case without parabolic fit. The problem of the smaller radius of the newly generated vertices without parabolic fit is again clearly shown in the negative radius error (Table \ref{SplittingOnlyResults}) for the simulation without parabolic fit. The absolute value of the radius error is more than two orders of magnitude lower if the new vertices are positioned based on the proposed parabolic fit. Because of the pronounced undulated front, the average curvature error without parabolic fit is more than four hundred times larger in comparison to the front obtained with the parabolic fit. The standard deviation of the curvature provides an indication of how uniform the curvature is over the entire front of the sphere. In the case of an ideal sphere, the curvature is the same at every point on the front, so the standard deviation is zero. Applying the parabolic fit, the standard deviation is two orders of magnitude smaller than without the parabolic fit, from which it can be concluded that shape of the front resulting from the parabolic fit is closer to an ideal sphere. \textcolor{black}{Furthermore, for the radius error, the curvature error as well as for the standard deviation of the curvature, the parabolic fit vertex positioning outperforms the interpolation in the barycentric coordinates by two orders of magnitude.}

In the second test scenario, with the same conditions as in the previous test, edge collapsing is now also applied as a mesh coarsening operation, in addition to the previously considered mesh refinement by means of edge splitting. The sphere is again expanded in radial direction until a radius of $R=0.3$ is reached, where edge splitting with or without parabolic fit, \textcolor{black}{or with the interpolation in the barycentric coordinates, is applied}. During the subsequent contraction phase to the initial size, the mesh is coarsened by edge collapsing with \textcolor{black}{one of the respective vertex positioning methods} to the approximate initial refinement level. The resulting errors for this test scenario are listed in Table \ref{Table:splittingandcollapsing}. In this case, the volume error with the parabolic fit is almost 28 times smaller than in the simulation without parabolic fit \textcolor{black}{and approximately half of the volume error of the interpolation in the barycentric coordinates}. Collapsing without parabolic fit increases the volume error by more than one order of magnitude, as can also be seen in the comparison of the two values of edge splitting only without parabolic fit (Table \ref{SplittingOnlyResults}) and edge splitting plus collapsing without parabolic fit (Table \ref{Table:splittingandcollapsing}). The radius and curvature error as well as the standard deviation of curvature hardly change in splitting and collapsing with parabolic fit compared to splitting only with parabolic fit. The parabolic fit vertex repositioning for the edge collapsing operation conserves the shape of the mesh as well as the parabolic fit does in vertex repositioning in the edge splitting operation. Edge collapsing without parabolic fit reduces the undulations previously caused by splitting without parabolic fit, but the resulting errors are still larger than for the front where the parabolic fit is applied for vertex repositioning. \textcolor{black}{Compared to the interpolation in the barycentric coordinates, the parabolic fit vertex positioning for edge splitting and collapsing is approximately one order of magnitude more accurate for all considered errors.}
\begin{table}[ht]
    \centering
     \caption{Relative errors of volume, radius, curvature and standard deviation of the curvature for the test case with only edge splitting during mesh refinement of an expanding sphere. After expansion until $t=4 R/U$ from initial radius $0.1$ to $0.3$, the sphere is contracted to its initial radius at which the errors are evaluated.}
    \label{SplittingOnlyResults}
    \begin{tabular}{ccccc}
    \toprule
    End conditions & $\epsilon_V$ & $\epsilon_R$ & $\epsilon_{\kappa}$ & $\sigma_{\kappa}$ \\
    \midrule
    with parabolic fit & 7.65e-3 & 3.73e-5 & -4.23e-3  & 3.43e-3 \\
    \midrule
    without parabolic fit & -4.87e-3 & -4.52e-3 & -1.96 & 4.83e-1 \\
    \midrule
    \textcolor{black}{barycentric interpolation} & \textcolor{black}{2.64-3} & \textcolor{black}{-1.78e-3} & \textcolor{black}{-2.32e-1} & \textcolor{black}{3.38e-1}
    \\
    \bottomrule
    \end{tabular}
\end{table}

\begin{table}[ht]
    \centering
     \caption{Relative errors of volume, radius, curvature and standard deviation of the curvature for the test case with edge splitting and edge collapsing \textcolor{black}{with the respective vertex positioning method} for the expanding and shrinking sphere. After expansion until $t=4 R/U$ from initial radius $0.1$ to $0.3$ with mesh refinement, the sphere is contracted, accompanied by mesh coarsening to its initial radius at which the errors are evaluated.}
    \label{Table:splittingandcollapsing}
    \begin{tabular}{ccccc}
    \toprule
    End conditions & $\epsilon_V$ & $\epsilon_R$ & $\epsilon_{\kappa}$ & $\sigma_{\kappa}$ \\
    \midrule
    with parabolic fit & 8.50e-4 & 4.03e-5 & -2.84e-3  & 4.45e-3 \\
    \midrule
    without parabolic fit & -2.34e-2 & -7.95e-3 & 1.52e-2 & 9.97e-2 \\
    \midrule
    \textcolor{black}{barycentric interpolation} & \textcolor{black}{-1.69e-3} & \textcolor{black}{-7.48e-4} & \textcolor{black}{-1.21e-2} & \textcolor{black}{2.63e-1}
    \\
    \bottomrule
    \end{tabular}
\end{table}

\subsubsection{Interface deformation}
The second test case for remeshing \textcolor{black}{with the respective vertex positioning methods} is the nonlinear deformation velocity field of \citet{LeVeque1996}, which already has been introduced in section \ref{Analytic deformation velocity field}. The initial spherical interface is positioned at $\begin{bmatrix}0.35 & 0.35 & 0.35 \end{bmatrix}$ with radius $R=0.15$ inside a unit cube domain. The average edge length of the triangles of the initial sphere is also the desired edge length $l$ for the remeshing threshold, which is $l = 0.0226$, or in relation to the initial radius $l/R = 0.15$. The interface is then advected and deformed with the exact velocity at each front vertex with a fourth-order Runge-Kutta method with a time step of $\Delta t = 0.013 R/U$ and $T=3$. During the first period up to $T/2$, the front mesh is refined by edge splitting and during the return to the initial position, for $T/2 < t < T$, it is coarsened by edge collapsing to approximately the initial refinement level. At $T$, the volume and shape parameters are computed and the relative errors compared to the start position are evaluated. Every fifth time step, the remeshing criteria and thresholds are checked and, if necessary, the remeshing operations are executed. The results for splitting and collapsing \textcolor{black}{with the respective vertex positioning method} are given in Table \ref{levequetable}. The variant with parabolic fit has a one order of magnitude more accurate volume conservative behaviour \textcolor{black}{compared to the vertex positioning without parabolic fit and has a volume error about four times smaller compared to the interpolation in the barycentric coordinates}. The curvature error for the end condition with parabolic fit is almost three orders of magnitudes more accurate compared to the end condition without parabolic fit \textcolor{black}{and also compared to the end condition of the curvature error for the interpolation in the barycentric coordinates}. 
\begin{table}[t]
    \centering
     \caption{Relative errors of volume, radius, curvature and standard deviation of the curvature for the deforming interface test case with edge splitting and edge collapsing \textcolor{black}{with the respective vertex positioning method}. The initial spherical interface is deformed until $T/2$, accompanied by mesh refinement by edge splitting. Afterwards the deformation is reversed, accompanied by mesh coarsening by edge collapsing, until the initial position is reached at $T = 3$ at which the errors are evaluated.}
    \label{levequetable}
    \begin{tabular}{ccccc}
\toprule
End conditions & $\epsilon_V$ & $\epsilon_R$ & $\epsilon_{\kappa}$ & $\sigma_{\kappa}$ \\
\midrule
with parabolic fit & 1.57e-3 & 5.60e-4 & 9.19e-5  & 2.44e-2 \\
\midrule
without parabolic fit & -1.82e-2 & -6.50e-3 & -5.52e-2 & 6.01e-2 \\
\midrule
    \textcolor{black}{barycentric interpolation} & \textcolor{black}{7.19e-3} & \textcolor{black}{-4.98e-4} & \textcolor{black}{1.22e-2} & \textcolor{black}{1.52e-1}
    \\
\bottomrule
\end{tabular}
\end{table}

\begin{figure}[ht]
    \centering
\includegraphics[]{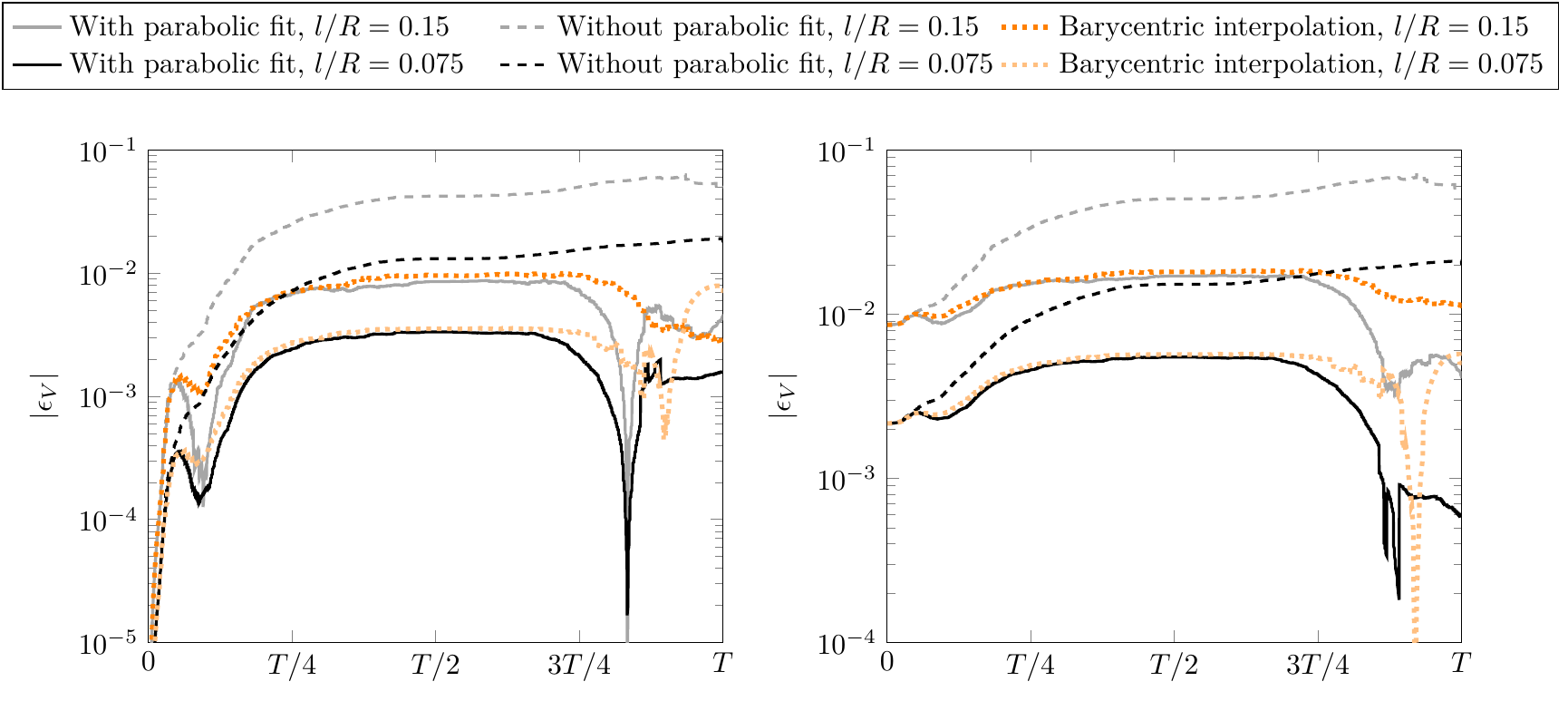}
    \caption{Relative volume errors over time for the deforming interface test case with edge splitting and collapsing. The initial spherical interface is deformed until $T/2$, accompanied by mesh refinement by edge splitting. Afterwards the deformation is reversed, accompanied by mesh coarsening by edge collapsing. The left diagram shows the volume error with the numerically calculated volume (equation (\ref{Volumenberechnung})) as reference value, and the right diagram shows the volume error with the analytical volume of an ideal sphere ($4 \pi R^3 / 3$) as the reference value.}
    \label{fig:VerrorOverTime}
\end{figure}

The velocity field used in this test case is divergence-free and the vertices of the triangles of the front are advected with the exact velocity in this case, so the enclosed volume should remain constant over time. However, the volume of the body still changes, due to the fact that it is only approximated by a triangle mesh. Nevertheless, next to shape preservation, a goal of remeshing is also to keep the volume error as low as possible over the entire course of the simulation. Figure \ref{fig:VerrorOverTime} shows the evolution of the volume error of the deforming front. The initial values are the same as before. In the left-hand diagram, the numerical volume at the start of the simulation calculated according to equation (\ref{Volumenberechnung}) serves as the reference value for the volume error, whereas in the right-hand diagram the volume of an ideal sphere calculated as $4 \pi R^3 /3$ serves as the reference value. Furthermore, two front mesh refinement levels are considered. Since the front vertices are advected with the exact velocities, fluid mesh spacing does not play a role. It can be seen that the volume error for remeshing without parabolic fit increases continuously for both refinement levels in both diagrams. In contrast, the volume errors for remeshing with parabolic fit \textcolor{black}{and the interpolation in the barycentric coordinates are} predominantly constant over time for both variants of volume error and about half an order of magnitude smaller than the error obtained without parabolic fit. The abrupt decrease of volume error for both front meshes with parabolic fit  \textcolor{black}{and the interpolation in the barycentric coordinates} shortly after $3T/4$ are related to a sign change. \textcolor{black}{However, after $3T/4$ a difference in accuracy between the parabolic fit and the interpolation in the barycentric coordinates becomes obvious and the parabolic fit leads to better results, as already shown in Table \ref{levequetable}.} Reducing the front mesh edge length by factor $2$ for remeshing with parabolic fit reduces the error compared to the numerically calculated starting volume by almost half an order of magnitude. For the error relative to the analytically calculated volume of an ideal sphere, reducing the edge length by factor $2$ leads to a reduction of the volume error by approximately one order of magnitude.
\begin{figure}[ht]
    \centering
\includegraphics[scale=0.9]{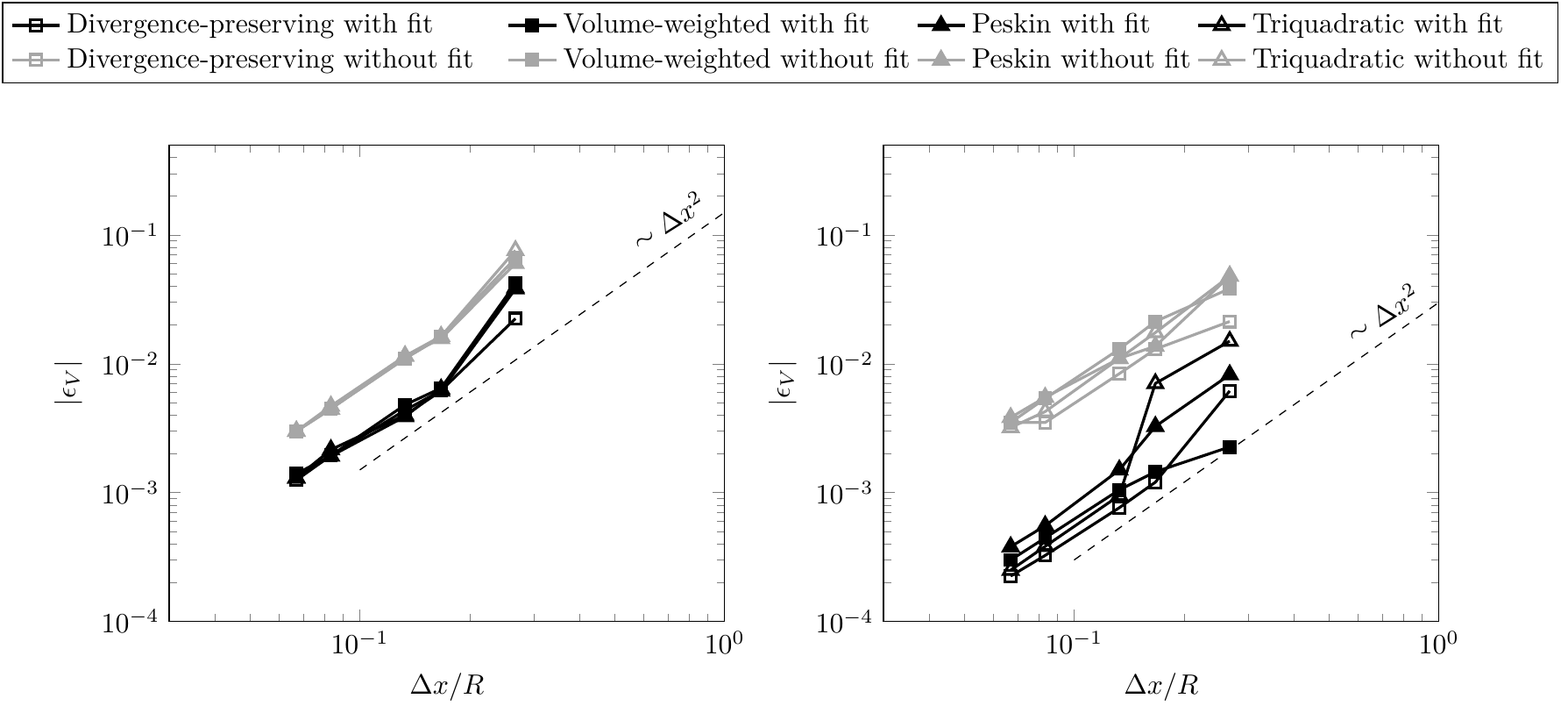}
    \caption{Relative volume errors for selected interpolation methods with remeshing with or without parabolic fit at various fluid mesh spacings $\Delta x$ for the deforming interface test case with edge splitting and collapsing. The initial spherical interface is deformed until $T/2$, accompanied by mesh refinement by edge splitting. Afterwards the deformation is reversed, accompanied by mesh coarsening by edge collapsing. The analytical volume of an ideal sphere ($4 \pi R^3 / 3$) serves as the reference value for the volume error. The left diagram shows the volume error at $T/2$ and the right diagram shows the volume error at $T$.}
    \label{fig:VerrorOverDeltaX}
\end{figure}

In order to assess the combined influence of remeshing with or without parabolic fit plus velocity interpolation on volume conservation, the volume error for simulations with velocity interpolation together with remeshing is considered in a final test series. Again, the same initial values as before are applied. As desired for front tracking simulations, in this test series, the desired and initial edge length $l$ for the front mesh as remeshing threshold is adapted to the fluid mesh spacing as $l=0.5\Delta x$. The CFL-number is kept constant at approximately $0.24$. In order to better compare the volume errors of the differently resolved front meshes, the analytically calculated volume of an ideal sphere ($4 \pi R^3 / 3$) is used as a reference value for the error calculation. The relative volume errors at $T/2$ and at $T$ for a selection of interpolation methods with remeshing with or without parabolic fit are shown in Figure \ref{fig:VerrorOverDeltaX}. At time $T/2$, the volume error is almost exclusively dominated by the error due to remeshing with or without parabolic fit. The curves of the selected interpolation methods all lie almost on top of each other, whereby the volume error is about three times larger for the variants without parabolic fit compared to the ones with parabolic fit. All variants converge with approximately second order when increasing the resolution of the fluid and front meshes. At the end of each simulation at $T$, the difference between the simulations with and without parabolic fit increases to nearly an order of magnitude smaller volume error for the simulations with parabolic fit. For the volume error at the end point, the velocity interpolation has an increased influence. For less resolved fluid and front meshes, the error due to velocity interpolation has a significant influence on the total volume error. This influence decreases with increasing mesh resolution, and for the highest resolutions considered, the influence of the interpolation methods on the volume error largely disappears and the error is dominated by remeshing with or without parabolic fit. As far as the velocity interpolation has an influence on the total volume error, the divergence-preserving velocity interpolation is consistently among the interpolation methods with the lowest volume error and converges with second order.

\section{Rising bubble}
\label{section:RisingBubble}
It is of great interest to test the combination of the presented methods not only for analytical velocity fields, but to couple the front tracking method with these two new attributes with a flow solver and to see if the presented methods still offer advantages. A rising bubble in a quiescent fluid is a classical example for validating front tracking methods. The bubble rises due to buoyancy effects resulting from the pressure gradient caused by gravity. The emerging velocity field of an incompressible flow around the bubble is divergence-free and, therefore, the bubble volume should remain constant.

The governing equations for this incompressible flow are the Navier-Stokes equations in the one-fluid formulation, which are given by the continuity equation
\begin{equation}
    \nabla \cdot \mathbf{u} = 0
\end{equation}
and the momentum equation
\begin{equation}
    \rho \left( \frac{\partial \mathbf{u}}{\partial t} + \nabla \cdot ( \mathbf{u} \otimes \mathbf{u}) \right) = - \nabla p + \nabla \cdot \tau + \rho \mathbf{g} + \mathbf{f}_{\sigma} \delta_S,
\end{equation}
where $\rho$ is the fluid density, $\mathbf{u}$ is the velocity vector, $p$ is the pressure, $\tau$ is the viscous stress tensor, $g$ is the gravity acceleration vector, $\mathbf{f}_{\sigma}$ is the surface tension force, and $\delta_S = \delta_S (\mathbf{x} - \mathbf{x}_S)$ is the interfacial delta function.
The incompressible Navier-Stokes equations are discretised and solved using a finite-volume framework with collocated variable arrangement, which solves the equations in a coupled, pressure-based manner with second order accuracy in space and time \cite{Denner2014a, Denner2020}.

The front tracking and the governing equations in the flow solver are coupled via the surface tension force, the indicator function for the material properties and the interpolation of the fluid velocity at the vertices of the triangles of the front. \textcolor{black}{At each time step, the interface is first advected from $t$ to $t+dt$, then the surface tension and indicator function are computed, and, subsequently, the Navier-Stokes equations are solved with the new interface position. It has to be mentioned that for the time instances $t+dt/2$ and $t+dt$ in the Runge-Kutta scheme applied for the front advection, the velocity field at time $t$ is used. If the velocity field changes rapidly in time, the accuracy of the scheme may, consequently, reduce to first order.}

The surface tension force is defined as
\begin{equation}
    \mathbf{f}_{\sigma} = \sigma \kappa \mathbf{n},
\end{equation}
with $\sigma$ the surface tension coefficient, $\kappa$ the curvature and $\mathbf{n}$ the normal vector of the interface. The surface tension force on a triangulated front element $E$ can be computed by using the Frenet-Element method given as  \cite{Tryggvason2001, Tryggvason2011a, Bi2021} 
\begin{equation}
    \mathbf{f}_{\sigma, E} = \sigma \int_{E} \kappa \mathbf{n} \mathrm{d}A = \sigma \oint_C \mathbf{p} \mathrm{d}l = \sigma \sum\limits_{e} \mathbf{p}_e \textcolor{black}{l_e}, 
\end{equation}
where with the help of the Stokes theorem the area integral of element $E$ with its boundary $C$ is converted into a line integral along the edges $e$ of the front triangle. The planar vector $\mathbf{p}_e = \mathbf{n}_e \times \mathbf{t}_e$ is perpendicular to edge $e$ and is given by the cross product of the normal vector of the front at edge $e$ and a tangential vector of the edge $e$. Afterwards, the surface tension force on all triangle front elements is spread onto the fluid mesh as
\begin{equation}
    \mathbf{f}_{\sigma}(\mathbf{x}) = \sum\limits_{E} \mathbf{f}_{\sigma, E} D(\mathbf{x} - \mathbf{x}_E), 
\end{equation}
where the distribution function $D(\mathbf{x} - \mathbf{x}_E)$ is in this work the Peskin weighting function, see equation (\ref{PeskinEquation})).

The reconstruction of the material properties $\phi(\mathbf{x},t)$, {\it i.e.} density and viscosity, at time $t$ on the fluid mesh is achieved by the use of an indicator function $I(\mathbf{x},t)$. This indicator function is set to zero in the liquid phase and one in the gas phase (the bubble). Therefore, the material properties are defined as
\begin{equation}
\phi(\mathbf{x},t) = \phi_l (1-I(\mathbf{x},t)) + \phi_g I(\mathbf{x},t),
\end{equation}
where the subscript $l$ indicates the liquid phase and $g$ the gas phase. The indicator function within the domain is constructed by solving the Poisson equation
\begin{equation}
    \nabla^2 I(\mathbf{x},t) = \nabla \cdot \left( \sum\limits_{E} \Delta I \mathbf{n}_E D(\mathbf{x} - \mathbf{x}_E) \right) ,
\end{equation}
where $\Delta I$ is the jump in the value of the indicator across the interface, $\mathbf{n}_E$ is the normal vector of front triangle element $E$ and, here, $D(\mathbf{x} - \mathbf{x}_E)$ represents again the Peskin function, see equation (\ref{PeskinEquation}). This indicator function, hence, varies smoothly in the interface region. A more detailed derivation of the indicator function can be found for instance in \cite{Tryggvason2011a, Tryggvason2001, Hua2008}.

In regions where the front area is shrinking, non-physical undulations can occur due to variations in the velocity field from fluid cell to fluid cell \cite{deSousa2004}. This phenomenon typically also occurs at the bottom side of rising bubbles simulated by the front tracking method and for this reason the volume conserving undulations removal algorithm TSUR3D by \citet{deSousa2004} is also implemented into the front remeshing procedure.

In this work, we consider a bubble rising in a stationary domain with periodic boundary conditions at the side walls, a no-slip bottom wall and an outlet at the top. The size of the domain is $8D\times8D\times20D$ \cite{Hua2008}, where $D$ is the initial bubble diameter. Nevertheless, the finite size of the domain still affects the rise velocity of the bubble. Following the work of \citet{Harmathy1960}, the expected rise velocity in an infinite domain can be approximated as
\begin{equation}
    U_{\infty} = \frac{U_T}{\left( 1 - \left( \frac{D}{L} \right)^2 \right)},
    \label{Harmathy}
\end{equation}
where $U_T$ is the terminal rise velocity in the finite domain, $D$ is the initial bubble diameter and $L$ is the domain size perpendicular to the rise axis. The non-dimensional parameters describing this flow test case are the bubble Reynolds number $\mathrm{Re} = \rho_l D U_T / \mu_l$, the Bond (or Eötvös) number $\mathrm{Bo} = \Delta \rho_l g D^2 / \sigma$, and the Morton number $\mathrm{Mo} = \Delta \rho_l g \mu^4_l / \rho_l^2 \sigma^3$. Furthermore, a dimensionless time $\tau = \sqrt{g/D} \cdot t$ is used. The liquid to gas density ratio is $1000$ and the viscosity ratio is $100$. The reference results for validation of the front tracking algorithm are taken from the experimental results from \citet{Bhaga1981}. Table \ref{ReynoldsNumberTest} shows the error of chosen velocity interpolation methods with or without parabolic fit for the terminal Reynolds number and the terminal Reynolds number for an infinite domain ($Re_\infty$ with the help of equation (\ref{Harmathy})) for $\tau = 15$ for a bubble with $\mathrm{Mo} = 266$ and \textcolor{black}{$\mathrm{Bo} = 243$}. All considered combinations of the velocity interpolation method with or without parabolic fit are in satisfying agreement with the experimental result from \citet{Bhaga1981}. The triquadratic interpolation without parabolic fit has the highest error in the terminal Reynolds number for the infinite domain with $1.0\%$, whereas the proposed divergence-preserving interpolation method with parabolic fit has the smallest error with $0.13\%$.

\begin{table}[ht]
    \centering
     \caption{Terminal Reynolds numbers and Reynolds number errors for chosen velocity interpolation methods with or without parabolic fit during remeshing operations at $\tau =15$, $\mathrm{Mo} = 266$ and \textcolor{black}{$\mathrm{Bo} = 243$}.}
    \label{ReynoldsNumberTest}
    \begin{tabular}{lcccc}
    \toprule
     & $Re$ & $|\epsilon_{Re}|$ & $Re_\infty$ &  $|\epsilon_{Re_\infty}|$\\
    \midrule
    Experiment \cite{Bhaga1981} & 7.77 &  \\
    \midrule
    Divergence-preserving with fit & 7.64 & 0.016 & 7.76 & 0.0013 \\
    \midrule
    Divergence-preserving without fit & 7.61 & 0.021 & 7.73 & 0.0051  \\
    \midrule
    Peskin with fit & 7.60 & 0.021 & 7.72 & 0.0064 \\
    \midrule
    Peskin without fit & 7.58 & 0.025 & 7.70 & 0.0090 \\
    \midrule
    Triquadratic with fit & 7.60 & 0.021 & 7.72 & 0.0064 \\
    \midrule
    Triquadratic without fit & 7.57 & 0.026 & 7.69 & 0.010 \\
    \bottomrule
    \end{tabular}
\end{table}

\begin{figure}[ht]
    \centering
\includegraphics[scale=0.9]{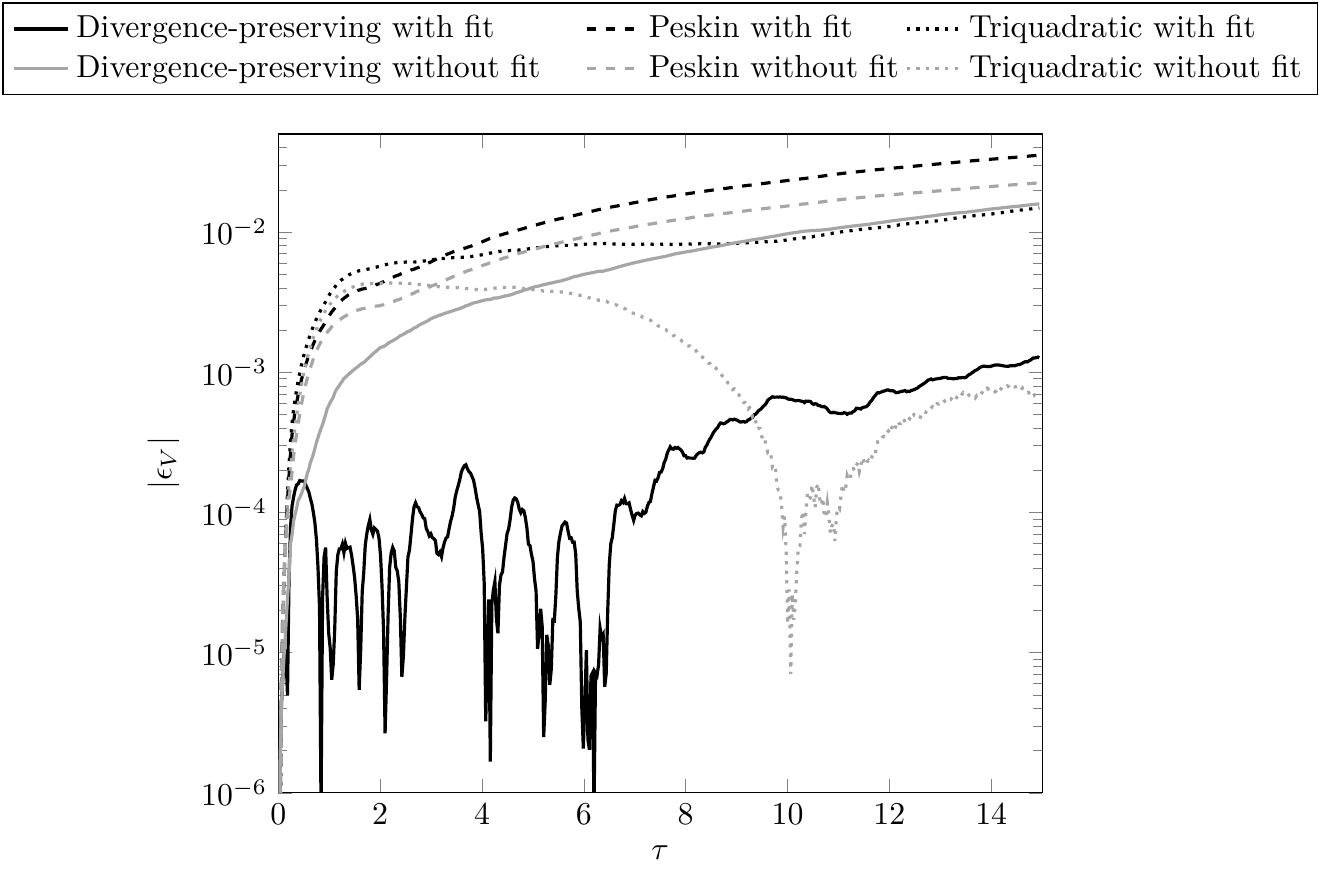}
    \caption{\textcolor{black}{Evolution of the volume error for selected interpolation methods with or without parabolic fit for a rising bubble at $\mathrm{Mo} = 266$ and} \textcolor{black}{$\mathrm{Bo} = 243$}.}
    \label{fig:VerrorRisingBubble}
\end{figure}

Typically, the volume error in front tracking simulations accumulates over time, which is why it is useful to consider not only the volume error during the acceleration phase of the rising bubble, but also when the terminal rise velocity has been reached. Figure \ref{fig:VerrorRisingBubble} shows the volume error over a long period of time for the rising bubble test case with $\mathrm{Mo} = 266$ and \textcolor{black}{$\mathrm{Bo} = 243$} for chosen velocity interpolation methods with or without the parabolic fit.
Figure \ref{fig:VerrorRisingBubble} shows that the volume error for all types of velocity interpolation methods with or without parabolic fit constantly increases over time. 
Solely the divergence-preserving interpolation method with parabolic fit fluctuates around the start volume and the maximum volume error can be kept nearly constant at approximately $10^{-4}$ up to $\tau \approx 7$. Afterwards the volume error slightly increases as well.
The volume error for the triquadratic interpolation method without parabolic fit decreases quite strongly around $8 \leq \tau \leq 10$, which is related to a sign change, probably arising from a favourable interplay between the remeshing and the TSUR3D algorithm. \textcolor{black}{Nevertheless, for the entire simulation, the proposed divergence-preserving velocity interpolation method with parabolic fit keeps the volume error small and has, together with the triquadratic interpolation method without fit, the smallest final volume error.}

\begin{figure}[ht]
    \centering
\includegraphics[scale=0.9]{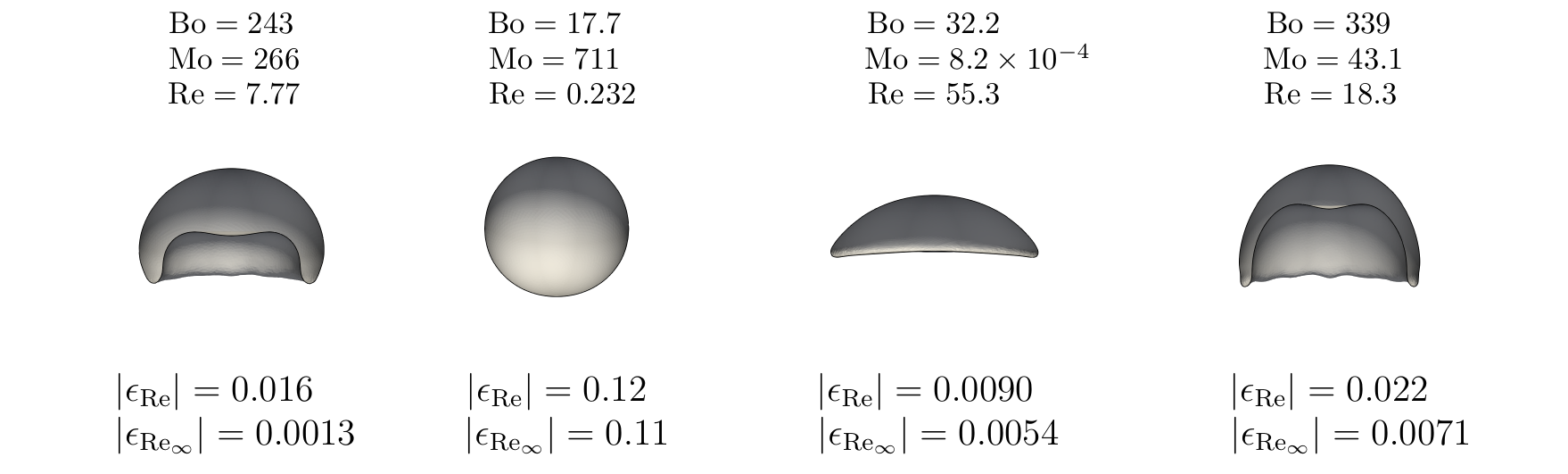}
    \caption{Bubble shapes and errors of the Reynolds numbers for the finite and infinite domain for a variety of Bond and Morten numbers.}
    \label{fig:RisingBubbleShapes}
\end{figure}

Furthermore, the presented divergence-preserving velocity interpolation method together with the parabolic fit was validated for a variety of Bond and Morten numbers. The terminal bubble shapes and Reynolds number errors are shown in Figure \ref{fig:RisingBubbleShapes}. For the error calculation, the numerically obtained Reynolds numbers are again compared with the experimentally determined Reynolds numbers from the work of  \citet{Bhaga1981}. All the bubble shapes and terminal Reynolds numbers are in good agreement with the experimental data and the errors are in the same order of magnitude or even smaller than, for instance, achieved in the front tracking work of \citet{Hua2008} or \citet{Pivello2014}. As already mentioned by \citet{Hua2008}, the higher error for the $\mathrm{Bo} = 17.7$, $\mathrm{Mo} = 711$, $\mathrm{Re} = 0.232$ test case most probably is due to the fact that the rise velocity is very low, wherefore even small absolute errors lead to high relative errors. Nevertheless, the error achieved in this test case is only half as big as the error ($0.21$) reported by \citet{Hua2008}.


\section{Summary and conclusions}
We have proposed a divergence-preserving velocity interpolation method and a parabolic fit vertex positioning method for front remeshing operations to increase the overall conservation of volume as well as accurately track the evolution of the interface in its advection with the front tracking method. The divergence-preserving interpolation method has been tested in a non-divergence-free and a divergence-free analytic velocity field, as well as for a rising bubble, and was compared with known interpolation methods commonly used in front tracking. It has been shown that the proposed method provides a higher accuracy in a classical finite-volume framework than the most commonly used trilinear (volume-weighted) and Peskin interpolation methods, and exhibits a similar or better accuracy than higher-order interpolation methods, such as the triquadratic interpolation method. Furthermore, the computational expense of the divergence-preserving method is comparable to the trilinear interpolation and converges with second order for increasing fluid mesh resolution.

Repositioning of the mid-vertex of a split or collapsed edge by a newly proposed parabolic fit has been shown to be superior in volume conservation as well as in shape and curvature preservation compared to the classically used mid-vertex position of the edge to be split or collapsed. While the volume error in remeshing without parabolic fit increases continuously during the course of the front advection, the volume error remains almost constant when the proposed parabolic fit is applied. Additionally, by construction, the parabolic fit preserves the shape and curvature of the local front, which is essential for an accurate computation of the geometric properties if surface tension is to be considered.

The volume conservation error resulting from the combination of velocity interpolation and remeshing is mainly dominated by the error generated by remeshing operations. For coarser fluid and front meshes, the error due to velocity interpolation has a higher influence on the total volume conservation error. This influence decreases with increasing mesh resolution and for the highest resolutions considered, the influence of interpolation on the volume error largely disappears. Overall, the divergence-preserving interpolation method together with remeshing with parabolic fit has consistently produced the most accurate and constant results for all fluid and front mesh resolutions considered in this study.

In the context of interfacial flow modelling, the proposed methods improve the overall accuracy and consistency of front tracking methods by addressing two elementary issues: volume conservation and shape preservation errors as results of an inaccurate representation and advection of the front mesh. The proposed divergence-preserving interpolation method in combination with a higher-order time integration scheme provides an accurate procedure to improve volume conservation and shape preservation during advection, at a similar computational expense as for the classically used velocity interpolation methods, such as trilinear (volume weighted) and Peskin interpolation. Additionally, the parabolic fit also ensures a consistently accurate representation of the interface throughout the remeshing processes, whereby an accurate calculation of the geometric properties is given throughout.

\section*{Acknowledgements}
This research was funded by the Deutsche Forschungsgemeinschaft (DFG, German Research Foundation), grant number 420239128. We would like to thank Marcio Pivello for the fruitful discussions on the remeshing algorithms.

\appendix

\section{Divergence-preserving velocity interpolation}
\label{Appendix A}
In this appendix we derive the correction terms for the divergence-preserving velocity interpolation, \textcolor{black}{originally introduced by \citet{Toth2002} for magnetohydrodynamics,} in the context of Lagrangian advection of the vertices of the triangles of the front.

For simplicity, a Cartesian 3D fluid mesh cell ranging from $\begin{bmatrix}-1 & -1 & -1 \end{bmatrix}$ to $\begin{bmatrix}1 & 1 & 1 \end{bmatrix}$ is considered. \textcolor{black}{As a reminder, $\bar{\mathbf{u}}^{\prime}$ is the linear interpolated velocity and $\bar{\mathbf{u}}$ is the interpolated velocity where the discrete divergence is preserved. $U,V,W$ are the face-centered velocities and the subscripts $x,y,z$ indicate the transverse gradients of the face-centered velocities, respectively.} To preserve the discrete divergence of the velocity field, corrections $f$, $g$, $h$ are introduced for each velocity component, such that inside the cell:
\begin{align}
    \bar{u}(x,y,z) & = \bar{u}^{\prime}(x,y,z) + f(x,y,z) \\
    \bar{v}(x,y,z) & = \bar{v}^{\prime}(x,y,z) + g(x,y,z) \\
    \bar{w}(x,y,z) & = \bar{w}^{\prime}(x,y,z) + h(x,y,z).
\end{align}
In addition, in order to guarantee continuity of the velocity component normal to the face, the correction for the velocity component normal to a face must vanish:
\begin{align}
    f(\pm1,y,z) & = 0 \label{Eq:BoundaryCondition1}\\
    g(x,\pm1,z) & = 0 \\
    h(x,y,\pm1) & = 0. \label{Eq:BoundaryCondition2}
\end{align}
The divergence of this corrected interpolated velocity, inside the cell, reads as
\begin{equation}
    \nabla \cdot \bar{\mathbf{u}} = \nabla \cdot \bar{\mathbf{u}}^{\prime} + f_x + g_y + h_z.
\end{equation}
Preserving the discrete divergence requires that
\begin{equation}
\begin{split}
        \nabla \cdot \bar{\mathbf{u}} - \nabla \cdot {\mathbf{U}} & = \frac{1}{2} \left[ \left(yU_y^{(1,0,0)} + zU_z^{(1,0,0)}\right) - \left(yU_y^{(-1,0,0)} + zU_z^{(-1,0,0)}\right) + 2 f_x\right.\\
     & \quad\quad\left. + \left(xV_x^{(0,1,0)} + zV_z^{(0,1,0)}\right) -  \left(xV_x^{(0,-1,0)} + zV_z^{(0,-1,0)}\right) + 2 g_y\right.\\
    & \quad\quad\left. + \left(xW_x^{(0,0,1)} + yW_y^{(0,0,1)}\right) -  \left(xW_x^{(0,0,-1)} + yW_y^{(0,0,-1)}\right) + 2h_z\right] = 0.
\label{eg:difference}
\end{split}
\end{equation}
\textcolor{black}{
To achieve this condition, the corrections $f$, $g$ and $h$ have to be second-order polynomials:  
\begin{align}
    f(x,y,z) & = A_f x^2 + B_f y^2 + C_f z^2 + D_f xy + E_f xz + F_f yz + G_f x + H_f y + I_f z + J_f \\
    g(x,y,z) & = A_g x^2 + B_g y^2 + C_g z^2 + D_g xy + E_g xz + F_g yz + G_g x + H_g y + I_g z + J_g\\
    h(x,y,z) & = A_h x^2 + B_h y^2 + C_h z^2 + D_h xy + E_h xz + F_h yz + G_h x + H_h y + I_h z + J_h
\end{align}
Using the boundary conditions from equations (\ref{Eq:BoundaryCondition1})-(\ref{Eq:BoundaryCondition2})  leads to:
\begin{align}
    f(1,y,z) - f(-1,y,z) & = 0 = 2D_f y + 2E_f z + 2G_f \quad \forall (y,z) \in [-1,1]^2 \\
    f(x,1,z) - f(x,-1,z) & = 0 = 2D_f x + 2F_f z + 2H_f \quad \forall (x,z) \in [-1,1]^2\\
    f(x,y,1) - f(x,y,-1) & = 0 = 2E_f x + 2F_f y + 2I_f \quad \forall (x,y) \in [-1,1]^2\\
    & \Rightarrow D_f = E_f = F_f = G_f = H_f = I_f = 0.
\end{align}
The same can be done for $g$ and $h$, so
\begin{align}
    f(x,y,z) & = A_f x^2 + B_f y^2 + C_f z^2 + J_f \\
    g(x,y,z) & = A_g x^2 + B_g y^2 + C_g z^2 + J_g\\
    h(x,y,z) & = A_h x^2 + B_h y^2 + C_h z^2 + J_h.
\end{align}
Using the boundary conditions again, we see that $f(\pm1,0,0) = 0$ so $A_f = -I_f$. Moreover, since $f(\pm1,y,z) = 0\ \forall (y,z) \in [-1,1]^2$, then $B_f = C_f = 0$. Therefore:
\begin{align}
    f(x,y,z) & = A_f (x^2-1) \label{eq:derivatives0}\\
    g(x,y,z) & = B_g (y^2-1) \\
    h(x,y,z) & = C_h (z^2-1), \label{eq:derivatives1}
\end{align}
which directly yields
\begin{align}
    -4A_f & = V_x^{(0,1,0)} - V_x^{(0,-1,0)} + W_x^{(0,0,1)} - W_x^{(0,0,-1)} \\
    -4B_g & = U_y^{(1,0,0)} - U_y^{(-1,0,0)} + W_y^{(0,0,1)} - W_y^{(0,0,-1)} \\
    -4C_h & = U_z^{(1,0,0)} - U_z^{(-1,0,0)} + V_z^{(0,1,0)} - V_z^{(0,-1,0)}.
\end{align}
As a result, the correction terms to preserve the exact discrete divergence inside the cell are:
\begin{align}
    f(x,y,z) & = \frac{1-x^2}{4} \left(V_x^{(0,1,0)} - V_x^{(0,-1,0)} + W_x^{(0,0,1)} - W_x^{(0,0,-1)}\right)\\
    g(x,y,z) & = \frac{1-y^2}{4} \left(U_y^{(1,0,0)} - U_y^{(-1,0,0)} + W_y^{(0,0,1)} - W_y^{(0,0,-1)}\right)\\
    h(x,y,z) & = \frac{1-z^2}{4} \left(U_z^{(1,0,0)} - U_z^{(-1,0,0)} + V_z^{(0,1,0)} - V_z^{(0,-1,0)}\right).
\end{align}
}

\end{document}